\documentclass[twocolumn,tighten]{aastex62}
\hypersetup{linkcolor=red,citecolor=blue,filecolor=cyan,urlcolor=magenta}
\usepackage{amsmath}
\usepackage{natbib}

\newcommand{\hdone}{\object[HD 184266]{HD~184266}}
\newcommand{\hdtwo}{\object[HD 222925]{HD~222925}}
\newcommand{\degree}{$^{\circ}$}
\newcommand{\loggf}{\mbox{$\log gf$}}
\newcommand{\kmsec}{\mbox{km~s$^{\rm -1}$}}
\newcommand{\logg}{\mbox{log~{\it g}}}
\newcommand{\msun}{\mbox{$M_{\odot}$}}
\newcommand{\teff}{\mbox{$T_{\rm eff}$}}
\newcommand{\vt}{\mbox{$v_{\rm t}$}}
\newcommand{\rpro}{\mbox{{\it r}-process}}
\newcommand{\spro}{\mbox{{\it s}-process}}

\newcommand{\ncap}{\mbox{{\it n}-capture}}
\newcommand{\rettwolong}{\object[NAME RETICULUM II]{Reticulum~II}}

\newcommand{\rtwo}{\mbox{\it r}-II}

\shorttitle{Abundances in HD 222925}
\shortauthors{Roederer et al.}
\accepted{for publication in the Astrophysical Journal}

\begin{document}

\title{%
The \textit{R}-Process Alliance:\ 
A Comprehensive Abundance Analysis of HD 222925, \\
a Metal-Poor Star with an Extreme \textit{r}-Process Enhancement 
of [Eu/H]~$= -0.14$\footnote{%
This paper includes data gathered with the 6.5~meter 
Magellan Telescopes located at Las Campanas Observatory, Chile.
}
}

\author{Ian U.\ Roederer}
\affiliation{%
Department of Astronomy, University of Michigan,
1085 S.\ University Ave., Ann Arbor, MI 48109, USA}
\affiliation{%
Joint Institute for Nuclear Astrophysics -- Center for the
Evolution of the Elements (JINA-CEE), USA}
\email{Email:\ iur@umich.edu}

\author{Charli M.\ Sakari}
\affiliation{%
Department of Astronomy, University of Washington, 
Seattle, WA 98195-1580, USA}

\author{Vinicius M.\ Placco}
\affiliation{%
Department of Physics, University of Notre Dame, 
Notre Dame, IN 46556, USA}
\affiliation{%
Joint Institute for Nuclear Astrophysics -- Center for the
Evolution of the Elements (JINA-CEE), USA}

\author{Timothy C.\ Beers}
\affiliation{%
Department of Physics, University of Notre Dame, 
Notre Dame, IN 46556, USA}
\affiliation{%
Joint Institute for Nuclear Astrophysics -- Center for the
Evolution of the Elements (JINA-CEE), USA}

\author{Rana Ezzeddine}
\affiliation{%
Department of Physics and Kavli Institute for Astrophysics and Space Research, 
Massachusetts Institute of Technology, 
Cambridge, MA 02139, USA}
\affiliation{%
Joint Institute for Nuclear Astrophysics -- Center for the
Evolution of the Elements (JINA-CEE), USA}

\author{Anna Frebel}
\affiliation{%
Department of Physics and Kavli Institute for Astrophysics and Space Research, 
Massachusetts Institute of Technology, 
Cambridge, MA 02139, USA}
\affiliation{%
Joint Institute for Nuclear Astrophysics -- Center for the
Evolution of the Elements (JINA-CEE), USA}

\author{Terese T.\ Hansen}
\affiliation{%
Carnegie Observatories,
Pasadena, CA 91101, USA}

\begin{abstract}

We present a detailed abundance analysis 
of the bright ($V =$~9.02),
metal-poor ([Fe/H]~$= -$1.47~$\pm$~0.08)
field red horizontal-branch star HD~222925,
which was observed as part of an ongoing survey by the
\textit{R}-Process Alliance.
We calculate stellar parameters and derive
abundances for 46~elements based on 901~lines examined in
a high-resolution optical spectrum obtained
using the Magellan Inamori Kyocera Echelle spectrograph.
We detect 28 elements with 38~$\leq Z \leq$~90;
their abundance pattern is a close match to 
the Solar \textit{r}-process component.
The distinguishing characteristic of HD~222925 is 
an extreme enhancement of \textit{r}-process elements
([Eu/Fe]~$= +$1.33~$\pm$~0.08,
[Ba/Eu]~$= -$0.78~$\pm$~0.10)
in a moderately metal-poor star, so
the abundance of \textit{r}-process elements
is the highest 
([Eu/H]~$= -$0.14~$\pm$~0.09)
in any known \textit{r}-process-enhanced star.
The abundance ratios among lighter ($Z \leq$~30) elements 
are typical for metal-poor stars,
indicating that production of these elements 
was dominated by normal Type~II supernovae,
with no discernible contributions from
Type~Ia supernovae or asymptotic giant branch stars.
The chemical and kinematic properties of HD~222925 
suggest it formed in a low-mass dwarf galaxy,
which was enriched by a high-yield \textit{r}-process 
event before being disrupted by interaction with the Milky Way.

\end{abstract}


\section{Introduction}
\label{intro}

The rapid neutron-capture process, or \rpro, 
is one of the fundamental ways to produce
the heaviest elements found in nature.
Decades of theoretical and observational efforts to
understand and characterize the \rpro\
are summarized in reviews by 
\citet{qian07}, \citet{sneden08}, \citet{thielemann17}, 
\citet{frebel18}, and \citet{horowitz18}.
Recent analysis of the
``kilonova'' electromagnetic counterpart 
(e.g., \citealt{cowperthwaite17,drout17,kasen17,tanvir17})
to a merger of
two neutron stars detected in gravitational waves
(GW170817; \citealt{abbott17prl,abbott17multimessenger})
confirms earlier observational suggestions
(e.g., \citealt{ji16nat,beniamini16b})
that neutron star mergers are viable \rpro\ sites.
No individual lines of \rpro\ elements can be
confidently identified in the kilonova spectra,
but hundreds of such lines are regularly 
detected in spectra of 
highly \rpro-enhanced stars in the Milky Way.
The existence, abundance patterns, and occurrence frequencies
of these stars
have established that the 
abundance pattern produced by the \rpro\
has remained largely unchanged 
across 9~Gyr of cosmic time before the
Sun was formed.

Highly \rpro-enhanced stars have minimal
contamination from the
slow neutron-capture process (\spro) and Eu/Fe ratios
$>$10 times higher than found in the Sun
(expressed as [Eu/Fe]~$> +$1.0,
where the abundance ratio of Eu and Fe relative to the
Solar ratio, [Eu/Fe], is defined as
$\log_{10} (N_{\rm Eu}/N_{\rm Fe}) - 
\log_{10} (N_{\rm Eu}/N_{\rm Fe})_{\odot}$).
These stars are often referred to as \rtwo\ stars
\citep{beers05}.
The abundance of the element Eu ($Z =$~63) is commonly used
to represent the level of \rpro\ enhancement.
\object[BPS CS 22892-052]{CS~22892-052},
which has [Eu/Fe]~$= +$1.6,
was the first recognized \rtwo\ star
\citep{sneden94}.
This star was identified
in the HK Survey of \citet{beers92}.
Since then
the average rate of discovery 
of \rtwo\ stars has been
$\lesssim$~1~yr$^{-1}$.
The low discovery rate reflects the rarity of these stars.
The occurrence frequency of \rtwo\ stars is $\approx$~3\%
among stars with [Fe/H]~$< -$1.5 \citep{barklem05heres}, 
which themselves only constitute
$\lesssim$~1\% of all stars in the Solar Neighborhood.

Expanding the sample of 
confirmed \rpro-enhanced stars
is one of the goals of a new effort called
the \textit{R}-Process Alliance (RPA).
The first large samples analyzed by the RPA
have been presented
by \citet{hansen18} and \citet{sakari18north},
and new discoveries of individual 
\rpro-enhanced stars have been presented by
\citet{placco17rpro}, 
\citet{cain18}, \citet{gull18},
\citet{holmbeck18rpro}, and \citet{sakari18a}.
Most RPA candidates have been selected from 
the RAdial Velocity Experiment 
\citep{kordopatis13rave,kunder17},
the LAMOST Survey \citep{liu14},
and the Best \& Brightest Survey
\citep{schlaufman14}.
Here, we present a new \rtwo\ star, \hdtwo.
We identified \hdtwo\ as a candidate for our observing program
by browsing the literature of the last few decades
in search of bright, metal-poor F- or G-type stars 
with insufficiently characterized heavy-element abundance patterns.

Table~\ref{datatab} lists the basic properties of \hdtwo.
\citet{houk75} classified it as a chemically peculiar ``Sr Eu'' star.
The abundances of metals in the atmosphere of \hdtwo\
reflect their bulk abundances in the star,
so it is not chemically peculiar in the
traditional sense.
\citeauthor{houk75} 
were, however, the first to recognize the presence of
strong second spectra lines
(i.e., lines arising from electronic 
transitions of ionized species) in \hdtwo.

Two modern studies have analyzed limited
sets of elements in \hdtwo.
\citet{gratton00} analyzed the pattern of light element abundance 
variations for 62 metal-poor stars at various stages of
stellar evolution.
They identified \hdtwo\ as a 
field equivalent of the cluster red horizontal-branch (RHB) stars,
deriving an effective temperature (\teff) of 5564~K,
surface gravity (\logg) of 2.64,
and metallicity ([Fe/H]) of $-$1.51.
\citet{navarette15} used \hdtwo\ 
as part of a control sample of field stars
in their abundance study that found no association between 
tidal stellar debris from $\omega$~Cen and the Kapteyn moving group.
\citeauthor{navarette15}\ derived \teff\ $=$ 5710~$\pm$~60~K,
\logg\ $=$ 2.32~$\pm$~0.14, and [Fe/H] $=$ $-$1.37~$\pm$~0.05.
Their study found that the He~\textsc{i} line at 10830~\AA\ 
in \hdtwo\ was several times stronger than that 
in other metal-poor field stars,
which they speculated could be due to chromospheric activity.
The three RHB stars in their study have the 
strongest He~\textsc{i} lines, so we 
presume that the line strength is related to the 
evolutionary state and not a natal enhancement of He.
Their study was also the first to quantify the
enhanced level of Ba in \hdtwo,
[Ba/Fe]~$= +$0.85~$\pm$~0.20,
but they did not consider any elements heavier than Ba.
\citeauthor{navarette15}\ suggested the enhanced Ba could result from
mass transfer of \spro\ rich material from an unseen companion star
that passed through the asymptotic giant branch (AGB)
phase of evolution.
They proposed to test this hypothesis by
searching for radial velocity (RV) variations 
or Y or Tc abundances, but
they did not pursue the matter further.

\begin{deluxetable*}{lcccc}
\tablecaption{Basic Data for HD 222925
\label{datatab}}
\tablewidth{0pt}
\tabletypesize{\scriptsize}
\tablehead{
\colhead{Quantity} &
\colhead{Symbol} &
\colhead{Value} &
\colhead{Units} &
\colhead{Reference} \\
}
\startdata
Right ascension           & $\alpha$ (J2000)    & 23:45:17.61            & hh:mm:ss.ss   & Simbad \\
Declination               & $\delta$ (J2000)    & $-$61:54:42.8          & dd:mm:ss.s    & Simbad \\
Galactic longitude        & $\ell$              & 316.0                  & degrees       & Simbad \\
Galactic latitude         & $b$                 & $-$53.5                & degrees       & Simbad \\
Parallax                  & $\varpi$            & 2.2332 $\pm$ 0.0243    & mas           & \citet{lindegren18} \\
Inverse parallax distance & $1/\varpi$          & 448 $\pm$ 5            & pc            & this study \\
Distance                  & $D$                 & 442$^{+4.9}_{-4.7}$    & pc            & \citet{bailerjones18} \\
Proper motion ($\alpha$)  & PMRA                & 154.854 $\pm$ 0.041    & mas yr$^{-1}$ & \citet{lindegren18} \\
Proper motion ($\delta$)  & PMDec               & $-$99.171 $\pm$ 0.041  & mas yr$^{-1}$ & \citet{lindegren18} \\
Radial velocity           & RV                  & $-$38.9 $\pm$ 0.6      & \kmsec        & this study \\
Mass                      & Mass                & 0.75 $\pm$ 0.20        & \msun         & assumed \\
$B$ magnitude             & $B$                 & 9.61 $\pm$ 0.02        & mag           & \citet{norris85} \\
$V$ magnitude             & $V$                 & 9.02 $\pm$ 0.02        & mag           & \citet{norris85} \\
$J$ magnitude             & $J$                 & 7.747 $\pm$ 0.023      & mag           & \citet{cutri03} \\
$H$ magnitude             & $H$                 & 7.415 $\pm$ 0.029      & mag           & \citet{cutri03} \\
$K$ magnitude             & $K$                 & 7.338 $\pm$ 0.026      & mag           & \citet{cutri03} \\
Color excess              & $E(B-V)$            & 0.00$^{+0.02}_{-0.00}$ & mag           & this study \\
Bolometric correction     & $BC_{V}$            & $-$0.21 $\pm$ 0.07     & mag           & based on \citet{casagrande14c} \\
Effective temperature     & \teff               & 5636 $\pm$ 103         & K             & this study \\
Log of surface gravity    & \logg               & 2.54 $\pm$ 0.17        & (cgs)         & this study \\
Microturbulent velocity   & \vt                 & 2.20 $\pm$ 0.20        & \kmsec        & this study \\
Model metallicity         & [M/H]               & $-$1.5 $\pm$ 0.1       & dex           & this study \\
Metallicitiy              & [Fe/H]              & $-$1.47 $\pm$ 0.08     & dex           & this study \\
\enddata      
%\tablecomments{}
%\tablenotetext{a}{}
\end{deluxetable*}

We present a comprehensive abundance analysis of \hdtwo\
based on new high-resolution optical spectroscopy.
Throughout this work,
we adopt the standard nomenclature 
for elemental abundances and ratios.
The absolute abundance of an element X is defined
as the number of X atoms per 10$^{12}$ H atoms,
$\log\varepsilon$(X)~$\equiv \log_{10}(N_{\rm X}/N_{\rm H})+$12.0.
%The abundance ratio of elements X and Y relative to the
%Solar ratio is defined as
% [X/Y] $\equiv \log_{10} (N_{\rm X}/N_{\rm Y}) - \log_{10} (N_{\rm X}/N_{\rm Y})_{\odot}$.
We adopt the Solar photospheric abundances of \citet{asplund09}.
By convention,
abundances or ratios denoted with the ionization state
are understood to be
the total elemental abundance, as derived from transitions of
that particular ionization state 
after Saha ionization corrections have been applied.

\section{Observations}
\label{obs}

We observed \hdtwo\ on 2017 September 26 using the
Magellan Inamori Kyocera Echelle 
spectrograph (MIKE; \citealt{bernstein03})
mounted at the $f/11$ focus on the east Nasmyth platform
of the Landon Clay (Magellan~II) Telescope at 
Las Campanas Observatory, Chile.
A pair of 50~s observations of \hdtwo\ using the 0\farcs7$\times$5\farcs0 
entrance slit and 2$\times$2 binning
revealed a strong Eu~\textsc{ii} absorption line
at 3819~\AA.~
Preliminary analyses over the subsequent 48 hours
indicated super-Solar [Eu/La] and [Eu/Ba] ratios, suggesting
that the heavy-element enhancement might be dominated by 
\rpro\ nucleosynthesis.

We reobserved \hdtwo\ with MIKE on 2017 September 28 
with a series of 300~s and 600~s exposures, totaling 115~min.
These observations used the
0\farcs35$\times$5\farcs0 entrance slit and the native 1$\times$1
detector binning.
This setup yielded a spectral resolving power 
$R \equiv \lambda/\Delta\lambda \sim$~68,000 on the blue spectrograph
($\lambda \lesssim$~5000~\AA) and $R \sim$~61,000 on the red spectrograph,
as measured from isolated emission lines in the comparison lamp spectra.
The outside temperature was changing somewhat during the observations,
and the instrument was slightly out of focus, so the
resolving power is lower than could otherwise be achieved with this
observing setup.
The observations were made with \hdtwo\ at an airmass between
1.19 and 1.22.
The seeing ranged from 0\farcs6 to 1\farcs0 arcsec, and thin but
variable clouds were present throughout these observations.
We also observed a bright, rapidly-rotating B3V star, 
\object[HIP 98412]{HIP~98412},
to divide out telluric lines from our spectra,
and a comparison metal-poor 
field RHB star, \hdone,
with the same MIKE setup.

We use the CarPy MIKE reduction pipeline \citep{kelson00,kelson03}
as the primary data reduction method.
This includes overscan subtraction, pixel-to-pixel flat field division,
image coaddition, cosmic ray removal, sky and scattered-light subtraction,
rectification of the tilted slit profiles along the orders,
spectrum extraction, and wavelength calibration.
We modify some of the default pipeline settings to 
work on data binned 1$\times$1 with the 0\farcs35$\times$5\farcs0 slit,
yet the wavelength solution produced by the pipeline
is unsatisfactory for the bluest
18 orders of our data.
For these, we manually generate the wavelength solution
using routines in the IRAF ``echelle'' package.
We also use IRAF to stitch together the individual orders,
continuum-normalize the spectra, and shift the
spectra to rest velocity.

Our final spectrum of \hdtwo\ covers 3330~$< \lambda <$~9410~\AA,
although the spectra longward of $\sim$~8000~\AA\ show
evidence of fringing.
Signal-to-noise (S/N) ratios in the continuum range from
$\sim$~80/1~pix$^{-1}$ near 3400~\AA,
$\sim$~250/1~pix$^{-1}$ near 4000~\AA,
$\sim$~500/1~pix$^{-1}$ near 4550~\AA,
$\sim$~500/1~pix$^{-1}$ near 5200~\AA,
to 
$\sim$~700/1~pix$^{-1}$ near 6750~\AA.
\hdtwo\ has fairly broad absorption lines
(median $\approx$~6.5~\kmsec\ for the Fe lines measured
in Section~\ref{felinelist}),
which is typical for RHB stars.
This results in fewer completely unblended lines,
but each line is greatly oversampled by 
$\approx$~20--30~pixels, or $\sim$~8 resolution elements, in our spectrum.
The predicted 3$\sigma$ line detection thresholds 
\citep{cayrel88,frebel08he}
for these data are $\approx$~1~m\AA.

\subsection{Radial Velocity}
\label{rvtext}

We measure the RV of \hdtwo\ by cross-correlating
the order containing the Mg~\textsc{i} \textit{b} triplet
against a metal-poor template observed with MIKE
(\object[HD~128279]{HD~128279}; see \citealt{roederer14d}).
We calculate the Heliocentric correction using the IRAF
``rvcorrect'' task.
Our measured Heliocentric RV,
$-$38.9~$\pm$~0.6~\kmsec,
agrees with that measured by \citet{navarette15},
$-$38.64~$\pm$~0.36~\kmsec,
and the RV reported by 
the second data release of the \textit{Gaia} mission
(DR2; \citealt{lindegren18,katz18}),
$-$37.93~$\pm$~0.28~\kmsec.
\citet{beers14} measured an RV of $-$34~$\pm$~7~\kmsec\ from
medium-resolution ($R \sim$~3000) spectroscopy,
which is consistent with these values.
Thus, \hdtwo\ shows no evidence of 
RV variations that would 
signal the presence of an unseen companion.

\section{Stellar Parameters}
\label{params}

\subsection{Fe Lines}
\label{felinelist}

We compile a list of Fe~\textsc{i} lines 
with reliable oscillator strengths from
the National Institute of Standards and Technology (NIST)
Atomic Spectral Database (ASD; \citealt{kramida18})
with grades B or better ($\leq$~10\% uncertainty, or 0.05~dex).
These \loggf\ values mainly come from work by
\citet{obrian91}.
We supplement this list with results from recent laboratory studies
with comparable uncertainties
\citep{denhartog14,ruffoni14,belmonte17}.
We discard Fe~\textsc{i} lines with lower excitation potential
(E.P.)\ $<$~1.2~eV, because previous studies have shown that
these lines may yield higher-than-average abundances 
in metal-poor dwarfs and giants
(e.g., \citealt{cayrel04,cohen08emp})
likely caused by departures from local thermodynamic equilibrium
(LTE) (e.g., \citealt{bergemann12}).
We also adopt \loggf\ values for
Fe~\textsc{ii} lines from NIST,
retaining lines with grades C or better
($\leq$~25\% uncertainty, 0.12~dex).

We measure equivalent widths (EWs) 
using a semi-automatic 
routine that fits Voigt or Gaussian line profiles to 
continuum-normalized spectra
\citep{roederer14c}.
We visually inspect each line, and
we discard from consideration any line that appears
blended, is subject to uncertain
continuum placement, or is otherwise compromised.
We examine a telluric spectrum simultaneously
with the stellar spectrum, and any lines
that appear to be contaminated with
telluric absorption are also discarded.
We restrict ourselves to lines with 
$\log$(EW/$\lambda$)~$< -$4.5.
We retain 124 Fe~\textsc{i} lines and
10 Fe~\textsc{ii} lines,
whose EWs are reported in Table~\ref{linetab}.

\begin{deluxetable}{cccccccc}
\tablecaption{Lines, Atomic Data, EWs, and Abundances
\label{linetab}}
\tablewidth{0pt}
\tabletypesize{\scriptsize}
\tablehead{
\colhead{Species} & 
\colhead{Wavelength} &
\colhead{E.P.} &
\colhead{\loggf} &
\colhead{\loggf} &
\colhead{EW} &
\colhead{Limit} &
\colhead{$\log\epsilon$} \\
\colhead{} &
\colhead{(\AA)} &
\colhead{(eV)} &
\colhead{} &
\colhead{ref.} &
\colhead{(m\AA)} &
\colhead{flag} &
\colhead{} 
}
\startdata
  Li\,I  & 6707.80 & 0.00 &  0.17 &  1 &\nodata &$<$&  0.800 \\ 
  Na\,I  & 4982.81 & 2.10 & -0.92 &  2 &    2.8 &   &  4.286 \\ 
  Na\,I  & 5682.63 & 2.10 & -0.71 &  2 &   10.5 &   &  4.581 \\ 
\enddata
\tablecomments{The complete version of Table~\ref{linetab} is available
in the online edition of the journal.
A sample is shown here to illustrate its form and content.}
\tablereferences{%
(1) \citet{smith98}, using HFS/IS from \citet{kurucz95};
(2) \citet{kramida18};
(3) \citet{pehlivanrhodin17};
(4) \citet{aldenius09};
(5) \citet{lawler89}, using HFS from \citet{kurucz95};
(6) \citet{lawler13};
(7) \citet{wood13};
(8) \citet{lawler14} for \loggf\ values and HFS;
(9) \citet{wood14v} for \loggf\ values and HFS, when available;
(10) \citet{sobeck07};
(11) \citet{lawler17};
(12) \citet{denhartog11} for \loggf\ values and HFS;
(13) \citet{obrian91};
(14) \citet{denhartog14};
(15) \citet{belmonte17};
(16) \citet{ruffoni14};
(17) \citet{lawler15} for \loggf\ values and HFS;
(18) \citet{wood14ni};
(19) \citet{kramida18}, using HFS/IS from \citet{kurucz95};
(20) \citet{roederer12b};
(21) \citet{morton00};
(22) \citet{biemont11};
(23) \citet{ljung06};
(24) \citet{nilsson08};
(25) \citet{wickliffe94};
(26) \citet{duquette85};
(27) \citet{hansen12} for \loggf\ value and HFS/IS;
(28) \citet{kramida18}, using HFS/IS from \citet{mcwilliam98} when available;
(29) \citet{lawler01la}, using HFS from \citet{ivans06} when available;
(30) \citet{lawler09};
(31) \citet{li07}, using HFS from \citet{sneden09};
(32) \citet{ivarsson01}, using HFS from \citet{sneden09};
(33) \citet{denhartog03}, using HFS/IS from \citet{roederer08a} when available;
(34) \citet{lawler06}, using HFS/IS from \citet{roederer08a} when available;
(35) \citet{lawler01eu}, using HFS/IS from \citet{ivans06};
(36) \citet{denhartog06};
(37) \citet{lawler01tb}, using HFS from \citet{lawler01tbhfs};
(38) \citet{wickliffe00};
(39) \citet{lawler04}, using HFS from \citet{sneden09};
(40) \citet{lawler08};
(41) \citet{wickliffe97tm}, using HFS from \citet{sneden09};
(42) \citet{sneden09} for \loggf\ value and HFS/IS;
(43) \citet{lawler09} for \loggf\ values and HFS;
(44) \citet{lawler07};
(45) \citet{quinet06};
(46) \citet{xu07}, using HFS/IS from \citet{cowan05};
(47) \citet{biemont00}, using HFS/IS from \citet{roederer12d};
(48) \citet{nilsson02th};
(49) \citet{nilsson02u}.
}
%\tablenotetext{a}{}
\end{deluxetable}

\subsection{Model Atmosphere Parameters}
\label{atmpars}

Table~\ref{datatab} summarizes the broadband photometry we have
compiled for \hdtwo.
The \citet{schlafly11} dust maps predict that the 
total Galactic reddening along the line of sight to \hdtwo\
is small, $E(B-V) =$~0.019.
We independently check the reddening by inspecting our 
spectrum for evidence of interstellar absorption
near the Na~\textsc{i} doublet at 5889 and 5895~\AA.~
We use the IRAF ``telluric'' task to 
remove telluric lines from this region of the spectrum of \hdtwo\ 
by comparing with our hot star standard.
No interstellar Na~\textsc{i} absorption is detected toward \hdtwo,
so we adopt $E(B-V) =$~0.00$^{+0.02}_{-0.00}$.
We de-redden using the 
extinction coefficients of \citet{mccall04}.

We calculate \teff\ from the metallicity-dependent
color-\teff\ relations presented by \citet{casagrande10}.
The zeropoint of this scale was determined 
using Solar twins, and 
\citet{casagrande14b} showed that it is 
also applicable to giants.
We adopt a metallicity of [Fe/H]~$= -$1.5~$\pm$~0.3
based on previous work \citep{gratton00,navarette15}.
We draw $10^{4}$ 
samples from each input parameter
(magnitudes, reddening, and metallicity)
and calculate the \teff\ value predicted by each one.
Each calculation is self-consistent and uses
the same set of input draws,
and we adopt
the median of the final distribution as \teff.
Five colors ($B-V$, $V-J$, $V-H$, $V-K$, $J-K$)
yield consistent estimates of
5644~$\pm$~133~K,
5628~$\pm$~74~K,
5641~$\pm$~64~K,
5630~$\pm$~54~K, and
5714~$\pm$~215~K, respectively.
The weighted average and statistical uncertainty in \teff\
is 5636~$\pm$~46~K.
We estimate the systematic uncertainty 
by performing the same set of calculations for two
other color-\teff\ calibrations presented by
\citet{alonso99b} and \citet{ramirez05b}.
These scales predict 5505~$\pm$~72~K and
5458~$\pm$~38~K, respectively.
We adopt the 
quadrature sum of the
statistical uncertainty from \citet{casagrande10} (46~K) and the
standard deviation of these three \teff\ values (92~K)
as the total uncertainty on \teff\ (103~K).

We calculate the \logg\ value 
from fundamental relations:\
\begin{eqnarray}
\log g = 4 \log \teff + \log(M/\msun) - 10.61 + 0.4(BC_{V}
  \nonumber\\
  + V + 5\log \varpi + 5 - 3.1 E(B-V) - M_{\rm bol,\odot}).
\end{eqnarray}
The symbols and their values 
are given in Table~\ref{datatab}.
$M_{\rm bol,\odot}$ is the Solar bolometric magnitude, 4.75, and
the constant 10.61 is calculated from the Solar constants
$\log \teff_{\odot} =$~3.7617 and $\log g_{\odot} =$~4.438.
We draw
$10^{4}$ samples from each
of these input parameters.
The median of these calculations gives the \logg\ value,
and their standard deviation gives the uncertainty:\
2.54~$\pm$~0.17.
The \teff\ and \logg\ values we calculate for \hdtwo\
are in good agreement with those determined by
\citet{gratton00} and \citet{navarette15},
quoted in Section~\ref{intro}.
They are also consistent with the values derived from
high-S/N medium-resolution spectroscopy by \citet{beers14},
\teff~$=$~5603~$\pm$~125~K and \logg~$=$~2.2~$\pm$~0.4.
% BM-325

We interpolate a one-dimensional,
hydrostatic model atmosphere from the $\alpha$-enhanced
ATLAS9 grid of models \citep{castelli04},
using an interpolation code provided by
A.\ McWilliam (2009, private communication).  
We derive Fe abundances 
using a recent version of the 
line analysis software MOOG
(\citealt{sneden73}; 2017 version).
MOOG assumes that LTE
holds in the line-forming layers of the atmosphere.
This version of the code treats 
Rayleigh scattering, which affects 
the continuous opacity at shorter wavelengths,
as isotropic, coherent scattering,
as described in \citet{sobeck11}.
We adopt 
damping constants for collisional broadening
with neutral hydrogen from \citet{barklem00h}
and \citet{barklem05feii}, when available,
otherwise
we adopt the standard \citet{unsold55} recipe.

We iteratively determine 
the microturbulent velocity, \vt, and model metallicity, [M/H].~
Lines yielding an abundance more than 
0.4~dex from the mean are culled.
Convergence is reached when there is no dependence
between line strength and abundance derived from
Fe~\textsc{i} lines, and when [M/H] equals the
average of the abundances derived from
Fe~\textsc{i} and Fe~\textsc{ii} lines,
rounded to the nearest 0.1~dex.
We find \vt~$=$~2.20~$\pm$~0.2~\kmsec\ and 
[M/H]~$= -$1.5~$\pm$~0.1~dex.
Our adopted model atmosphere parameters for \hdtwo\
are listed in Table~\ref{datatab}.

We compute [Fe~\textsc{i}/H] and [Fe~\textsc{ii}/H] ratios,
and their difference, by drawing 
$10^{3}$ samples from each input parameter in the model atmosphere
(\teff, \logg, \vt, and [M/H]) from normal distributions.
We interpolate a new model atmosphere 
for each of these draws, 
and the abundances are recomputed for each line.
We associate an EW uncertainty with each line, given by
$\sqrt{(0.05\times{\rm EW})^{2} + 1.0}$,
which asymptotes to 1~m\AA\ for the weakest lines and 
5\% for the strongest lines.
We sample 
the \loggf\ value for each line
from a normal distribution
whose dispersion is given by the \loggf\ uncertainty
(see references to Table~\ref{linetab}).
We adopt the median of these $10^{3}$ realizations
as the average Fe abundance.
The 16th and 84th percentiles of the distributions
are roughly symmetric, so we report 
one number 
as the systematic uncertainty, $\sigma$.
These values are reported in Table~\ref{abundtab}.

We find only a small offset between
the Fe abundances derived
from Fe~\textsc{i} and Fe~\textsc{ii} lines,
[Fe~\textsc{i}/H]~$= -$1.58~$\pm$~0.01 ($\sigma =$~0.08~dex) and
[Fe~\textsc{ii}/H]~$= -$1.47~$\pm$~0.03 ($\sigma =$~0.08~dex).
Their difference,
[Fe~\textsc{ii}/H]$-$[Fe~\textsc{i}/H]
$= +$0.11~$\pm$~0.03~dex ($\sigma =$~0.10~dex),
is small but significant.
This suggests that transitions in neutral Fe may not be
adequately characterized by Boltzmann and Saha LTE calculations,
even when low-E.P.\ Fe~\textsc{i} lines and strong lines 
are excluded from consideration.
Non-LTE overionization may be responsible.
Singly-ionized Fe atoms are expected to dominate ($>$~98\%)
by number in the line-forming layers of \hdtwo,
so LTE is an acceptable approximation for Fe~\textsc{ii} lines.
We confirm this hypothesis by
interpolating non-LTE corrections for 14 Fe~\textsc{i} 
lines in common with the INSPECT database
\citep{bergemann12,lind12}.
Their average non-LTE correction is $+$0.12~dex,
which would bring the [Fe/H] ratio derived from Fe~\textsc{i} lines
into good agreement with that derived from Fe~\textsc{ii} lines.
Extrapolating the simple linear relation between [Fe/H] and
the non-LTE correction to abundances derived from Fe~\textsc{i} lines
found by \citet{ezzeddine17} predicts a similar correction 
of $+$0.07~dex.
We conclude that departures from LTE are likely
responsible for the discrepancy between
the [Fe/H] ratios derived from Fe~\textsc{i} and \textsc{ii} lines
in \hdtwo.

\begin{deluxetable}{lccccc}
\tablecaption{Derived Abundances 
\label{abundtab}}
\tablewidth{0pt}
\tabletypesize{\scriptsize}
\tablehead{
\colhead{Species} &
\colhead{$\log\varepsilon$} &
\colhead{$\sigma$} &
\colhead{[X/Fe]\tablenotemark{a}} &
\colhead{$\sigma$} &
\colhead{$N_{\rm lines}$} \\
}
\startdata
Fe~\textsc{i}   &    5.92 &  0.08 &    $-$1.58 &  0.08 &124  \\
Fe~\textsc{ii}  &    6.03 &  0.08 &    $-$1.47 &  0.08 & 10  \\
Li~\textsc{i}   & $<$0.80 &\nodata&    \nodata &\nodata&  1  \\
C (CH)          &    6.65 &  0.15 &    $-$0.20 &  0.17 &\nodata \\
N (NH)          &    6.45 &  0.20 &    $+$0.20 &  0.21 &\nodata \\
Na~\textsc{i}   &    4.49 &  0.07 &    $-$0.17 &  0.07 &  7  \\
Mg~\textsc{i}   &    6.43 &  0.08 &    $+$0.41 &  0.05 &  7  \\
Al~\textsc{i}   &    4.80 &  0.17 &    $-$0.07 &  0.17 &  3  \\
Si~\textsc{i}   &    6.38 &  0.07 &    $+$0.45 &  0.07 & 18  \\
K~\textsc{i}    &    3.68 &  0.10 &    $+$0.23 &  0.04 &  1  \\
Ca~\textsc{i}   &    5.12 &  0.08 &    $+$0.36 &  0.05 & 34  \\
Sc~\textsc{ii}  &    1.82 &  0.13 &    $+$0.14 &  0.08 &  8  \\
Ti~\textsc{i}   &    3.60 &  0.11 &    $+$0.23 &  0.03 & 13  \\
Ti~\textsc{ii}  &    3.88 &  0.09 &    $+$0.40 &  0.05 & 14  \\
V~\textsc{i}    &    2.38 &  0.11 &    $+$0.03 &  0.04 &  3  \\
V~\textsc{ii}   &    2.65 &  0.11 &    $+$0.19 &  0.08 & 12  \\
Cr~\textsc{i}   &    3.95 &  0.11 &    $-$0.11 &  0.03 & 15  \\
Cr~\textsc{ii}  &    4.16 &  0.08 &    $-$0.01 &  0.05 &  6  \\
Mn~\textsc{i}   &    3.55 &  0.08 &    $-$0.30 &  0.02 & 13  \\
Mn~\textsc{ii}  &    3.78 &  0.19 &    $-$0.18 &  0.16 &  3  \\
Co~\textsc{i}   &    3.33 &  0.18 &    $-$0.08 &  0.10 & 21  \\
Ni~\textsc{i}   &    4.62 &  0.09 &    $-$0.02 &  0.02 & 13  \\
Cu~\textsc{i}   &    2.05 &  0.15 &    $-$0.56 &  0.12 &  1  \\
Zn~\textsc{i}   &    3.17 &  0.06 &    $+$0.19 &  0.04 &  3  \\
Rb~\textsc{i}   & $<$2.10 &\nodata&   $<+$1.16 &\nodata&  2  \\
Sr~\textsc{i}   &    1.46 &  0.11 &    $+$0.17 &  0.05 &  1  \\
Sr~\textsc{ii}  &    1.98 &  0.13 &    $+$0.58 &  0.13 &  3  \\
Y~\textsc{ii}   &    1.04 &  0.10 &    $+$0.30 &  0.07 & 40  \\
Zr~\textsc{ii}  &    1.74 &  0.10 &    $+$0.63 &  0.07 & 51  \\
Nb~\textsc{ii}  &    0.61 &  0.16 &    $+$0.62 &  0.15 &  1  \\
Mo~\textsc{i}   &    1.30 &  0.15 &    $+$1.00 &  0.09 &  3  \\
Ru~\textsc{i}   &    1.34 &  0.14 &    $+$1.17 &  0.08 &  5  \\
Rh~\textsc{i}   &    0.64 &  0.16 &    $+$1.31 &  0.12 &  3  \\
Pd~\textsc{i}   &    1.05 &  0.15 &    $+$1.06 &  0.08 &  3  \\
Ag~\textsc{i}   &    0.44 &  0.18 &    $+$1.08 &  0.13 &  1  \\
Ba~\textsc{ii}  &    1.26 &  0.09 &    $+$0.55 &  0.06 &  5  \\
La~\textsc{ii}  &    0.51 &  0.09 &    $+$0.88 &  0.07 & 40  \\
Ce~\textsc{ii}  &    0.85 &  0.08 &    $+$0.74 &  0.07 & 67  \\
Pr~\textsc{ii}  &    0.22 &  0.10 &    $+$0.97 &  0.08 & 22  \\
Nd~\textsc{ii}  &    0.88 &  0.09 &    $+$0.93 &  0.08 & 99  \\
Sm~\textsc{ii}  &    0.62 &  0.09 &    $+$1.13 &  0.08 & 87  \\
Eu~\textsc{ii}  &    0.38 &  0.09 &    $+$1.33 &  0.08 & 17  \\
Gd~\textsc{ii}  &    0.82 &  0.09 &    $+$1.22 &  0.08 & 38  \\
Tb~\textsc{ii}  &    0.16 &  0.10 &    $+$1.33 &  0.08 &  3  \\
Dy~\textsc{ii}  &    1.01 &  0.11 &    $+$1.38 &  0.08 & 32  \\
Ho~\textsc{ii}  &    0.12 &  0.15 &    $+$1.11 &  0.12 &  9  \\
Er~\textsc{ii}  &    0.74 &  0.10 &    $+$1.29 &  0.08 & 13  \\
Tm~\textsc{ii}  & $-$0.13 &  0.10 &    $+$1.24 &  0.08 &  7  \\
Yb~\textsc{ii}  &    0.48 &  0.18 &    $+$1.11 &  0.17 &  1  \\
Lu~\textsc{ii}  &    0.06 &  0.13 &    $+$1.43 &  0.09 &  2  \\
Hf~\textsc{ii}  &    0.37 &  0.09 &    $+$0.99 &  0.09 &  5  \\
Os~\textsc{i}   &    1.26 &  0.15 &    $+$1.44 &  0.09 &  2  \\
Ir~\textsc{i}   &    1.54 &  0.16 &    $+$1.74 &  0.11 &  1  \\
Pb~\textsc{i}   & $<$1.10 &\nodata&   $<+$0.93 &\nodata&  1  \\
Th~\textsc{ii}  & $-$0.06 &  0.12 &    $+$1.39 &  0.11 &  5  \\
U~\textsc{ii}   &$<-$0.50 &\nodata&   $<+$1.51 &\nodata&  2  \\
\enddata      
%\tablecomments{}
\tablenotetext{a}{[Fe/H] values listed for Fe~\textsc{i} and Fe~\textsc{ii}.}
\end{deluxetable}

% totals:
%
% 901 lines total
% 571 lines of n-cap elements
%
% 55 species, including upper limits
% 51 species detected
%
% 50 elements, including upper limits (plus H)
% 46 elements detected (plus H)
%
% 31 n-capture elements, including upper limits
% 28 n-capture elements detected
%
% 19 lighter elements, including upper limits (plus H)
% 18 lighter elements detected (plus H)

\section{Visual Inspection of the Spectrum of HD 222925}
\label{inspection}

\begin{figure*}
\begin{center}
\includegraphics[angle=0,width=5.0in]{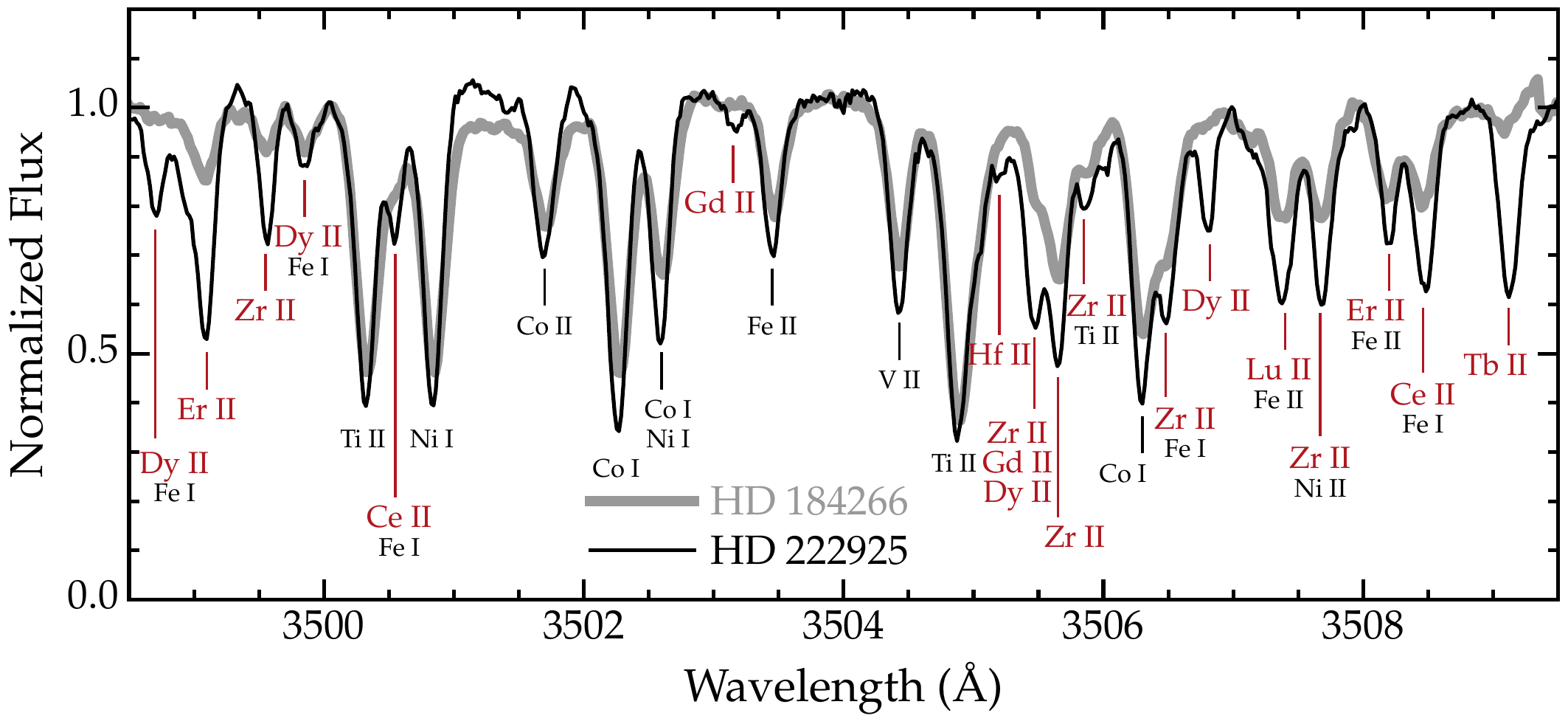} \\
\vspace*{0.1in}
\includegraphics[angle=0,width=5.0in]{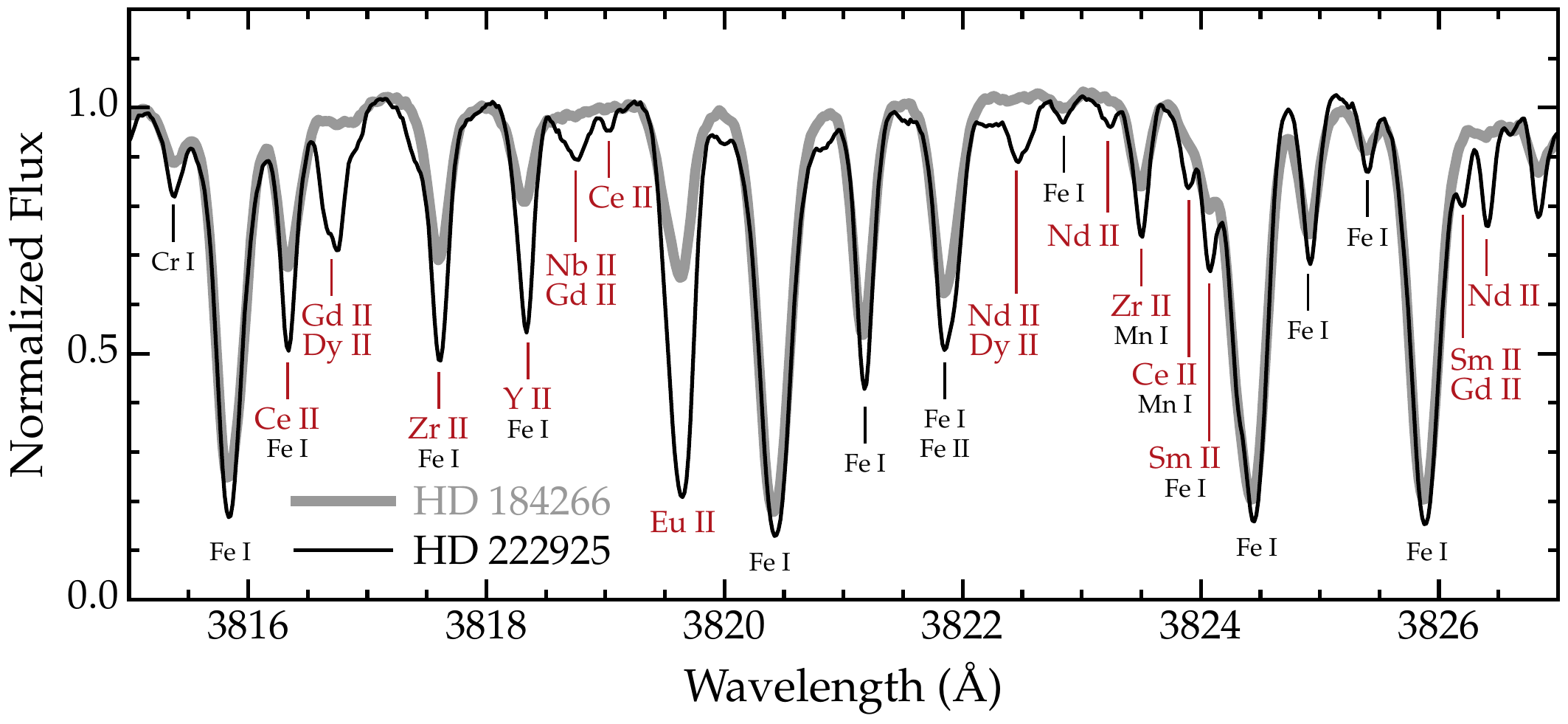} \\
\vspace*{0.1in}
\includegraphics[angle=0,width=5.0in]{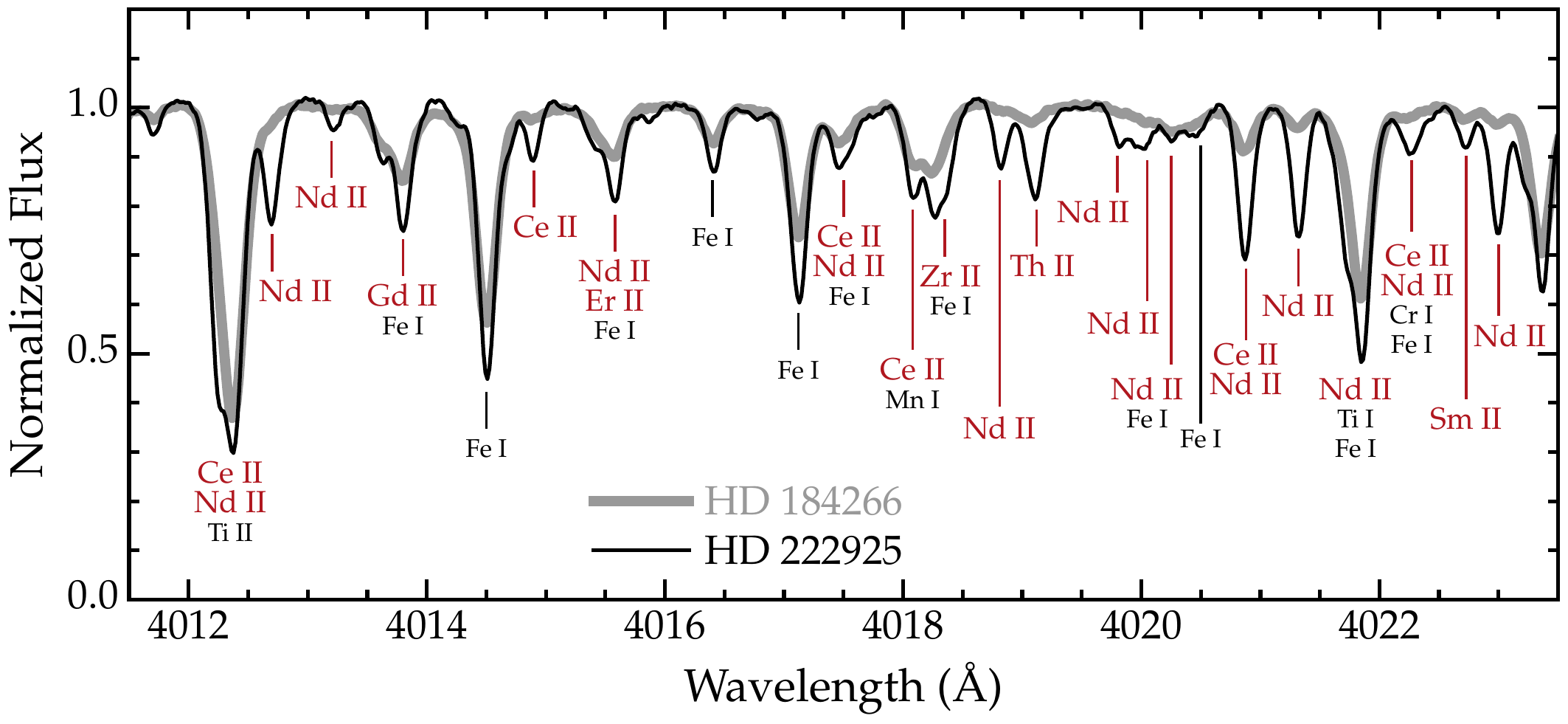} \\
\end{center}
\caption{
\label{specplots}
Comparison of sections of the spectra of \hdtwo\ and \hdone.
Lines of heavy ($Z >$~30) elements are labeled in red,
and lines of lighter elements are labeled in black.
The spectrum of \hdone\ was taken using the same MIKE setup.
\hdtwo\ and \hdone\ have similar stellar parameters,
but the abundances of heavy elements
are considerably lower in \hdone\
([Ba/Fe]~$= +$0.05, [Eu/Fe]~$= +$0.29; \citealt{roederer14c}).
 }
\end{figure*}

Figure~\ref{specplots} illustrates three small regions of the
spectrum of \hdtwo.
Another RHB star with similar stellar parameters, \hdone,
is shown for comparison
in Figure~\ref{specplots}.
We rederive the stellar parameters for \hdone\ using the
methods described in Section~\ref{atmpars}.
We find 
\teff~$=$~5580~$\pm$~102~K,
\logg~$=$~2.50~$\pm$~0.19,
\vt~$=$~2.25~$\pm$~0.20~\kmsec,
and
[Fe~\textsc{i}/H]~$= -$1.80~$\pm$~0.08,
which are similar to the values we find for \hdtwo,
\teff~$=$~5636~$\pm$~103~K,
\logg~$=$~2.54~$\pm$~0.17,
\vt~$=$~2.20~$\pm$~0.20~\kmsec, 
and
[Fe~\textsc{i}/H]~$= -$1.58~$\pm$~0.08.
It is apparent from Figure~\ref{specplots} that the 
lines of Fe-group species in \hdone\ are slightly weaker
than those in \hdtwo, and our analysis confirms
that the metallicity of \hdone\ is slightly lower.
The contrast between the two stars is 
most apparent in the strengths of absorption lines 
from heavy ($Z >$~30) elements, which are 
indicated in red.
Many of these lines are weak or absent in the 
spectrum of \hdone, but they are strong in the spectrum of \hdtwo.

Figure~\ref{specplots} also demonstrates
how incredibly rich the spectrum of \hdtwo\ is 
with lines of heavy elements.
Fourteen species of heavy elements are detectable 
in these three spectral windows.
The other remarkable characteristic of the spectrum of \hdtwo\
is the contrast between the lines of Fe-group elements
and heavy elements.
The abundance of Eu atoms in \hdtwo\ is 72\% of that in the Sun
([Eu/H]~$= -$0.14; calculated from data in Table~\ref{abundtab}),
yet the abundance of Fe atoms in \hdtwo\ is only 3\% of that in the Sun
([Fe/H]~$= -$1.47).
The warm, low-pressure atmosphere of \hdtwo\
further minimizes blends from lines 
of neutral Fe-group elements
while enhancing lines of ionized \ncap\ elements.

\section{Abundance Analysis}
\label{abundances}

We use the MOOG ``abfind'' driver
to derive abundances of most elements
with $Z \leq$~30 
based on EW measurements.
These values are reported in Table~\ref{linetab}.
Lines of 
Sc~\textsc{ii}, V~\textsc{i} and \textsc{ii},
Mn~\textsc{i} and \textsc{ii}, Co~\textsc{i},\footnote{%
Ionized Co, the majority species, is not 
often analyzed in the spectra of late-type stars.
Close inspection of Figure~\ref{specplots} reveals
a relatively unblended 
Co~\textsc{ii} line at 3501.72~\AA\ in \hdtwo.
A recent laboratory analysis of Co~\textsc{ii} \loggf\ values
by \citet{lawler18}
did not report a \loggf\ value for this line,
because it is a weak branch from the upper level
and is blended in their spectrum.
The six other Co~\textsc{ii} lines with \loggf\ values
reported by \citeauthor{lawler18}\ that are 
covered by our MIKE spectrum appear
blended in \hdtwo,
and we are unable to derive Co abundances from them.}
and Cu~\textsc{i}
are broadened by hyperfine splitting structure (HFS),
so we derive their abundances 
by spectrum synthesis matching using the MOOG ``synth'' driver.
All elements heavier than Zn are also derived by
spectrum synthesis matching.
We also derive an upper limit on the
Li, Rb, Pb, and U abundances using 
spectrum synthesis matching.
We derive abundances or upper limits 
from 901~lines in the spectrum on \hdtwo,
including 571~lines of elements with $Z >$~30.
Table~\ref{linetab} lists the wavelengths of these lines,
their E.P.\ values, \loggf\ values, references for the
\loggf\ values and any HFS or isotope shifts (IS)
considered in the syntheses, and the derived abundances.
Multiple isotopes are considered in the synthesis
for Li, C, Cu, Ag, Ba, Nd, Sm, Eu, Yb, Ir, and Pb.
We adopt $^{7}$Li/$^{6}$Li~$=$~1000,
$^{12}$C/$^{13}$C~$=$~5 (see below),
$^{63}$Cu/$^{65}$Cu~$=$~2.24 (the Solar ratio),
and the \rpro\ isotopic fractions from \citet{sneden08}
for all other elements.

We derive C and N abundances by iteratively
fitting portions of the
CH \textit{G}-band (4290--4315~\AA) and the
NH band (3355--3365~\AA).
We estimate [C/Fe]~$= -$0.20~$\pm$~0.17 and
[N/Fe]~$= +$0.20~$\pm$~0.21,
giving C/N~$=$~1.6.
These molecular features are relatively weak
in the spectrum, and no $^{13}$CH features
are detected with confidence,
so we simply adopt $^{12}$C/$^{13}$C~$=$~5
in our syntheses.
Carbon is depleted during the normal course of stellar evolution,
and the natal abundance in \hdtwo\ may have been higher
by $\approx +$0.46~dex \citep{placco14c},
yielding an initial [C/Fe]~$\approx +$0.26.
\hdtwo\ was never C-enhanced according to standard definitions
\citep{aoki07cemp}.

We cannot reliably measure the O abundance from the
[O~\textsc{i}] line at 6300.30~\AA.~
This line is intrinsically weak, and it 
is also blended with a telluric feature.
\citet{navarette15} derived a normal [O/Fe] ratio
for \hdtwo, $+$0.42~$\pm$~0.07, 
based on the \citet{ramirez07} non-LTE corrections
to the the O~\textsc{i} triplet
at 7771, 7773, and 7774~\AA.
Our EW measurements for these lines
are in good agreement with the \citeauthor{navarette15}\ values,
so we simply adopt their [O/Fe] ratio.

We derive uncertainties on the $\log\varepsilon$ 
abundances and [X/Fe] ratios 
(where ``X'' represents different elements)
using a Monte Carlo approach.
We sample the model atmosphere parameters,
line EWs (or approximations of
the EWs for lines examined by spectrum synthesis matching), 
and \loggf\ values $10^{3}$ times,
and rerun MOOG for each of these samples.
This technique implicitly captures the covariances
between element ratios, and
these samples are run simultaneously with the
[Fe/H] samples, so the results are self-consistent.
Table~\ref{abundtab} lists the median abundances, uncertainties,
and number of lines used to derive the abundance
for each species.
The 16th and 84th percentiles of the distributions are
roughly symmetric, so only one number is listed
as the uncertainty
for each abundance or ratio in Table~\ref{abundtab}.

Non-LTE corrections for Na~\textsc{i} lines,
when available from the INSPECT database
\citep{lind11},
are reflected in the values 
presented in Tables~\ref{linetab} and \ref{abundtab}.
The average non-LTE correction for six of the Na~\textsc{i} lines
is $-$0.16~dex.
The Na~\textsc{i} line at 4982.81~\AA\ is not
included in the INSPECT database,
so we adopt its LTE abundance.
We adopt the non-LTE correction, $-$0.62~dex,
interpolated from the grid of \citet{takeda02}
for the K~\textsc{i} line at 7698.96~\AA.
This correction is also reflected in the values
presented in Tables~\ref{linetab} and \ref{abundtab}.

The [X/Fe] ratios derived from the minority (neutral) species
of Ti, V, and Mn are slightly deficient (0.10--0.16~dex)
relative to the ratios derived from their ions.
Similarly, the [Sr/Fe] ratio derived from the
Sr~\textsc{i} line at 4607.33~\AA\ is 
deficient by 0.41~dex relative to that
derived from Sr~\textsc{ii} lines.
Overionization is likely affecting these species
(cf.\ \citealt{bergemann08,bergemann11,hansen13sr}),
so we favor the abundances and ratios
derived from lines of their ions.
The [Cu/Fe] ratio, derived from one Cu~\textsc{i} line,
likely underestimates the true value by a few tenths of a dex
\citep{korotin18,roederer18a}.
Most heavy elements are detected in ionization states
that represent a substantial fraction, if not
a majority, of all atoms of each element.
Departures from LTE are expected to be small
for these species.
Pb is an exception.  
We derive an upper limit from
one Pb~\textsc{i} line, and
\citet{mashonkina12} have shown that non-LTE
corrections can be substantial 
($>$~0.2~dex) for Pb~\textsc{i} lines
in metal-poor dwarf and giant stars.
It is possible that the upper limit we have inferred is underestimated
by a few tenths of a dex in LTE.

\subsection{Abundances in the Balmer Dip Region}
\label{balmerdip}

Previous studies have shown that abundances of 
Ti~\textsc{i}, Ti~\textsc{ii}, and Fe~\textsc{i}
yield abundances that are lower by $\approx$~0.08--0.27~dex
when derived from lines in the 3100--3700~\AA\ region 
in warm metal-poor dwarfs
\citep{wood13,lawler13,sneden16,roederer18b}
and cool giants \citep{roederer12d}.
This phenomenon is referred to as the Balmer Dip effect.
These studies concluded that 
unaccounted continuous opacity, 
non-LTE effects in individual metal ions or levels, 
and non-LTE effects in the H~\textsc{i} $n =$~2 level
cannot fully explain all observations available at present.
Three-dimensional convection effects have been suggested as
a possible explanation.

Nine species in our study have at least three lines in the
Balmer Dip region
and at least three lines at longer wavelengths,
which we consider minimally sufficient to assess whether
a similar effect may occur in \hdtwo.
Three of these species are Fe-group species
(V~\textsc{ii}, Fe~\textsc{i}, and Co~\textsc{i}), and
the other six species are heavy elements
(Y~\textsc{ii}, Zr~\textsc{ii}, Gd~\textsc{ii}, 
Dy~\textsc{ii}, Ho~\textsc{ii}, and Er~\textsc{ii}).
The differences between mean abundances derived from lines
with $\lambda <$~3700~\AA\ and $\lambda >$~3700~\AA\
are not significant at the $\approx$~1.2$\sigma$ level.
Only Zr~\textsc{ii} shows a marginally significant
discrepancy, where 35 Zr~\textsc{ii} lines with 
$\lambda >$~3700~\AA\ yield an abundance 
0.10~$\pm$~0.04~dex higher than 16 Zr~\textsc{ii} lines
at shorter wavelengths.
\citet{roederer18b} noted that the effect appeared muted,
when it appeared at all in warmer stars.
Our results support and extend that conclusion.

\section{Discussion}
\label{discussion}

\subsection{Elements with $Z \leq$ 30}
\label{lightelements}

\begin{figure*}
\includegraphics[angle=0,width=2.25in]{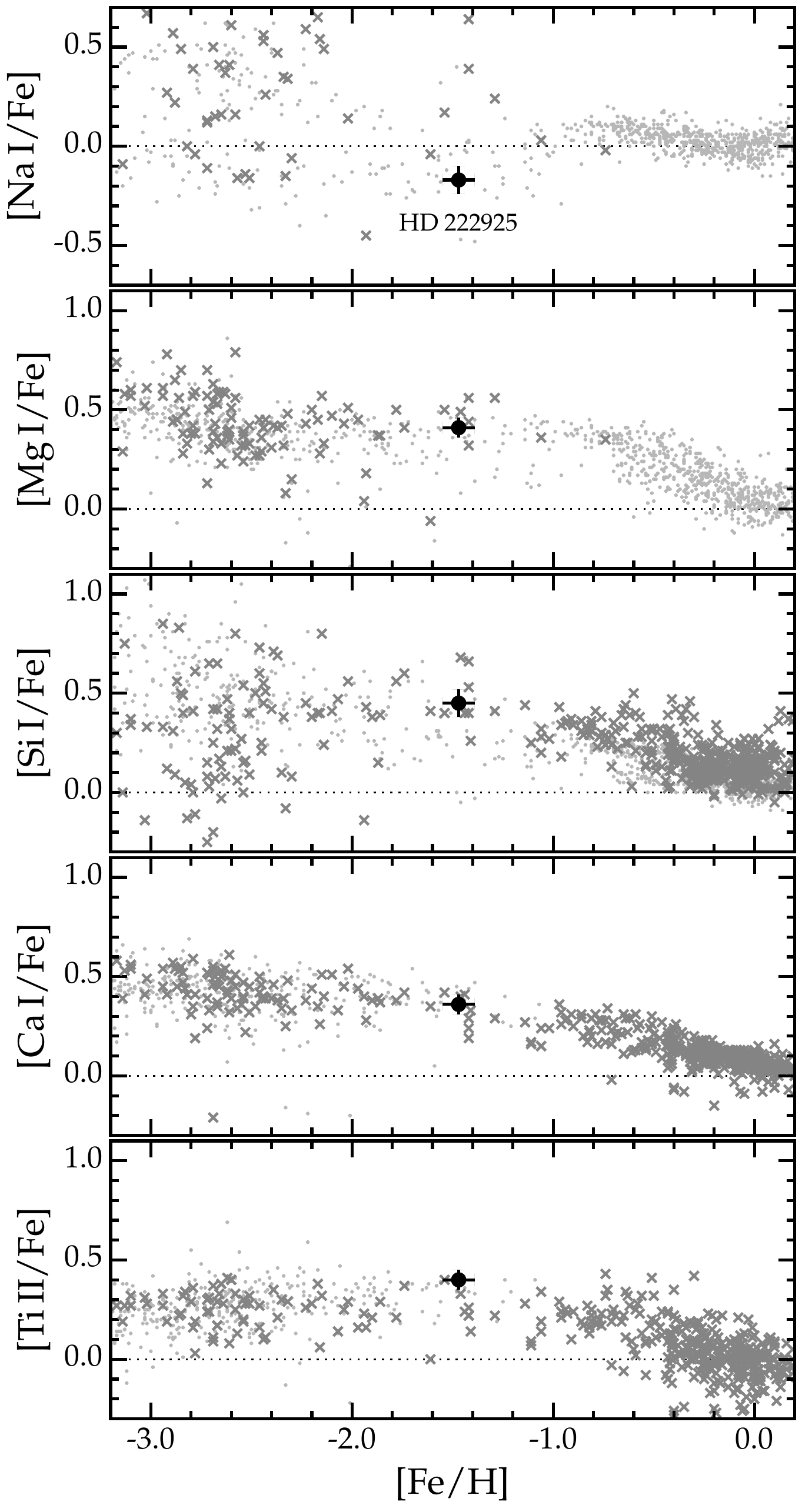} 
\hspace*{0.1in}
\includegraphics[angle=0,width=2.25in]{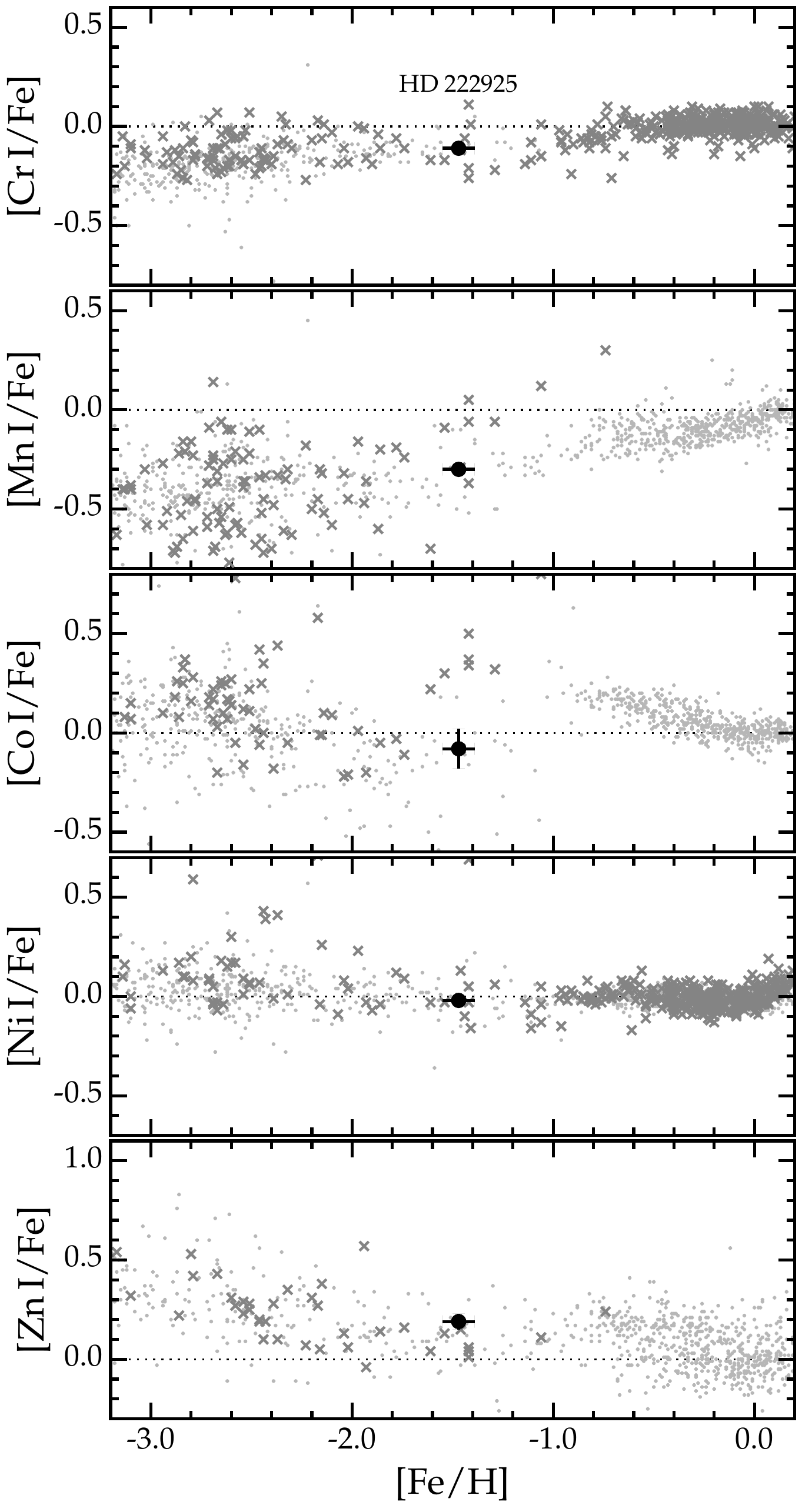} 
\hspace*{0.1in}
\includegraphics[angle=0,width=2.25in]{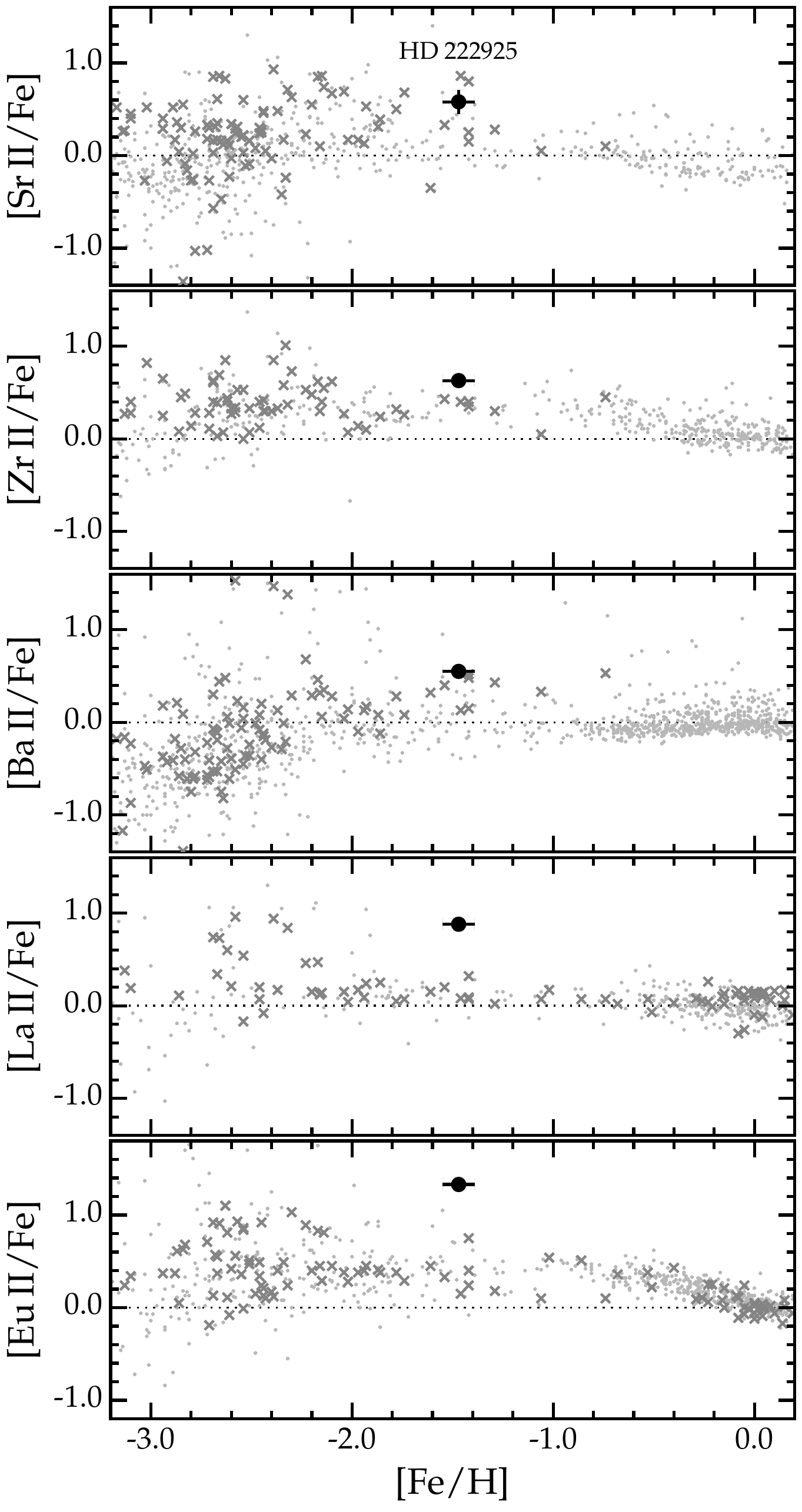} 
\caption{
\label{xfecomparisonplot}
Comparison of abundance ratios in \hdtwo\ (large black circle) 
with other field stars.
The small gray crosses represent 
RHB stars from 
\citet{preston06},
\citet{for10},
\citet{roederer14c}, and
\citet{afsar12,afsar18}.
The small gray dots represent dwarfs and giants from 
\citet{bensby14}, 
\citet{roederer14c},
\citet{jacobson15smss},
\citet{battistini15,battistini16}, and
\citet{hansen18}.
The dotted line in each panel represents the Solar ratio.
 }
\end{figure*}

We derive abundances for 18 metals with $Z \leq$~30 in \hdtwo.
Figure~\ref{xfecomparisonplot} illustrates the [X/Fe] ratios,
where X represents a particular element.
Several sets of abundance ratios from the literature are
shown for comparison.  
We prioritize comparisons with
analyses of large numbers of RHB stars in the field,
which may help to minimize systematic uncertainties.
Abundance ratios from the RHB star samples of 
\citet{preston06}, \citet{for10}, 
\citet{roederer14c}, and \citet{afsar12,afsar18}
are shown by dark gray crosses in Figure~\ref{xfecomparisonplot}.
These samples offer few stars for comparison in the 
metallicity range around [Fe/H]~$= -$1.5,
so we supplement with data from
other stellar types.
Abundance ratios from the dwarf and giant star samples of
\citet{bensby14}, \citet{roederer14c}, \citet{jacobson15smss},
\citet{battistini15,battistini16}, and \citet{hansen18} are
shown by small gray dots in Figure~\ref{xfecomparisonplot}.

Figure~\ref{xfecomparisonplot} demonstrates that 
the abundances of elements with $Z \leq$~30 in \hdtwo\
match those of other metal-poor field stars
with $-$2~$<$~[Fe/H]~$< -$1.
The $\alpha$-elements O (see Section~\ref{abundances}),
Mg, Si, Ca, and Ti
are enhanced relative to the Sun, with 
[$\alpha$/Fe]~$\approx +$0.4.
The [Na/Fe] ratio and [Al/Fe] ratio (not shown) 
are both slightly sub-Solar in \hdtwo,
which matches the comparison samples
(see also \citealt{andrievsky07,andrievsky08}).
The [K/Fe] ratio (not shown) is slightly super-Solar, 
and this is also normal for stars in this metallicity range
\citep{roederer14c}.
Among the Fe-group elements,
[Zn/Fe] is super-Solar,
[Sc/Fe] and [V/Fe] (both not shown) are slightly super-Solar,
[Cr/Fe], [Mn/Fe], [Co/Fe], and [Ni/Fe] are Solar or slightly sub-Solar,
and [Cu/Fe] (not shown) is significantly
sub-Solar in \hdtwo.
All of these ratios are normal for stars in this metallicity range.

Mixing processes may affect the 
surface composition of C and N in evolved stars 
like \hdtwo.
The [C/Fe] ratio in \hdtwo\ is slightly sub-Solar,
and we infer its natal abundance (Section~\ref{abundances})
to have been slightly super-Solar,
[C/Fe]~$= +$0.26, 
which does not qualify as being C enhanced.
The [N/Fe] ratio in \hdtwo\ is slightly super-Solar.
Both of these ratios are normal for
stars on the RHB \citep{gratton00}.

In summary, 
the agreement between the abundances in \hdtwo\ and the 
comparison samples implies that
normal Type~II supernovae produced most of the
metals with $Z \leq$~30.
There are minimal, if any, contributions from Type~Ia supernovae or
AGB stars.
\hdtwo\ formed in a region
where chemical evolution was dominated by massive stars
that enriched the local ISM to [Fe/H]~$= -$1.5 relatively quickly.

\subsection{Elements with $Z \geq 38$}
\label{rpropattern}

Figure~\ref{xfecomparisonplot}
illustrates the high levels of [Sr/Fe], [Zr/Fe],
[Ba/Fe], [La/Fe], and [Eu/Fe] in \hdtwo.
The [Sr/Fe], [Zr/Fe], and [Ba/Fe] ratios 
are found along the upper envelope of ratios
in the comparison samples, while
[La/Fe] and [Eu/Fe] are significantly higher.
Such high levels of enhancement are not generally
found among stars with [Fe/H]~$> -$2.
The high [Eu/Fe] ratio ($+$1.33~$\pm$~0.08)
and low [Ba/Eu] ratio ($-$0.78~$\pm$~0.10)
identify \hdtwo\ as a member of the 
\rtwo\ class of stars.

\begin{figure*}
\begin{center}
\includegraphics[angle=0,width=5.0in]{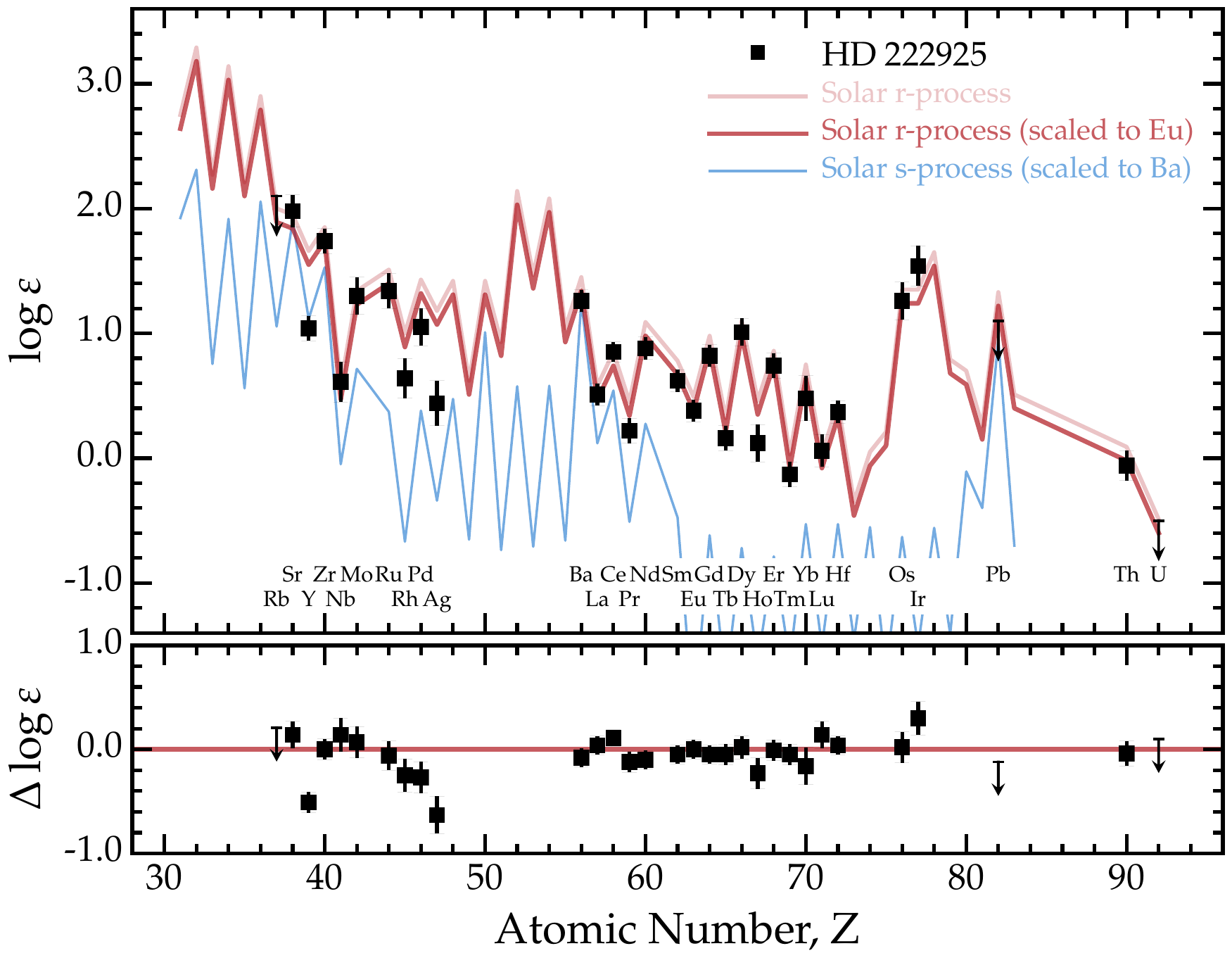}
\end{center}
\caption{
\label{rpropatternplot}
Top panel:\ 
Comparison between the heavy-element abundances in
\hdtwo\ and the Solar \textit{r-} and \spro\ patterns
\citep{sneden08}.
The Solar \rpro\ residual pattern (thick red line) is normalized
to the Eu abundance, and 
the Solar \spro\ pattern (thin blue line) is normalized 
to the Ba abundance.
The pink line marks an unscaled version of the Solar \rpro\ residual pattern.
Bottom panel:\
Differences between the \hdtwo\ abundances and the 
scaled Solar \rpro\ residuals.
 }
\end{figure*}

The heavy-element abundances in \hdtwo\
are illustrated in Figure~\ref{rpropatternplot}.
Three patterns are shown for comparison
in the top panel.
The pink line represents the Solar \rpro\ residuals
\citep{sneden08}; this line has not been rescaled.
The dark red line represents the Solar \rpro\ residuals
when scaled down by 0.11~dex to match the Eu abundance
in \hdtwo.
The thin blue line represents the Solar \spro\ abundance pattern
scaled down by 0.95~dex to match the Ba abundance in \hdtwo.

The Solar \rpro\ residuals fit most elements well.
The lanthanides (57~$\leq Z \leq$~71), plus Ba and Hf
($Z =$~56 and 72),
exhibit a robust \rpro\ abundance pattern.
This agreement extends to the third \rpro\ peak 
(Os and Ir; $Z =$~76 and 77).
There are no significant deviations 
between the overall levels of Sr and Zr
($Z =$~38 and 40)
and the scaled Solar \rpro\ residuals.
The \rpro\ event that enriched \hdtwo\
produced, at a minimum, substantial quantities
of elements from 38~$\leq Z \leq$~90.

It is also apparent from Figure~\ref{rpropatternplot}
that the Solar \spro\ pattern is a poor match to the 
\hdtwo\ abundances, regardless of how it is normalized.
Pb lies at the end of the \spro\ nucleosynthesis chain,
and it is often observed to be highly enhanced
([Pb/Fe]~$> +$2) in metal-poor stars with strong \spro\ signatures
(e.g., \citealt{aoki02pb,vaneck03}).
Pb is often the first detectable
signature of \spro\ contamination \citep{roederer10c}.
The reasonably low limit on Pb in \hdtwo,
with or without non-LTE corrections,
indicates that \spro\ material is no more than minimally present in \hdtwo.

\subsubsection{Deviations among Some Lighter \textit{R}-Process Elements}
\label{downturn}

\begin{figure}
\begin{center}
\includegraphics[angle=0,width=2.8in]{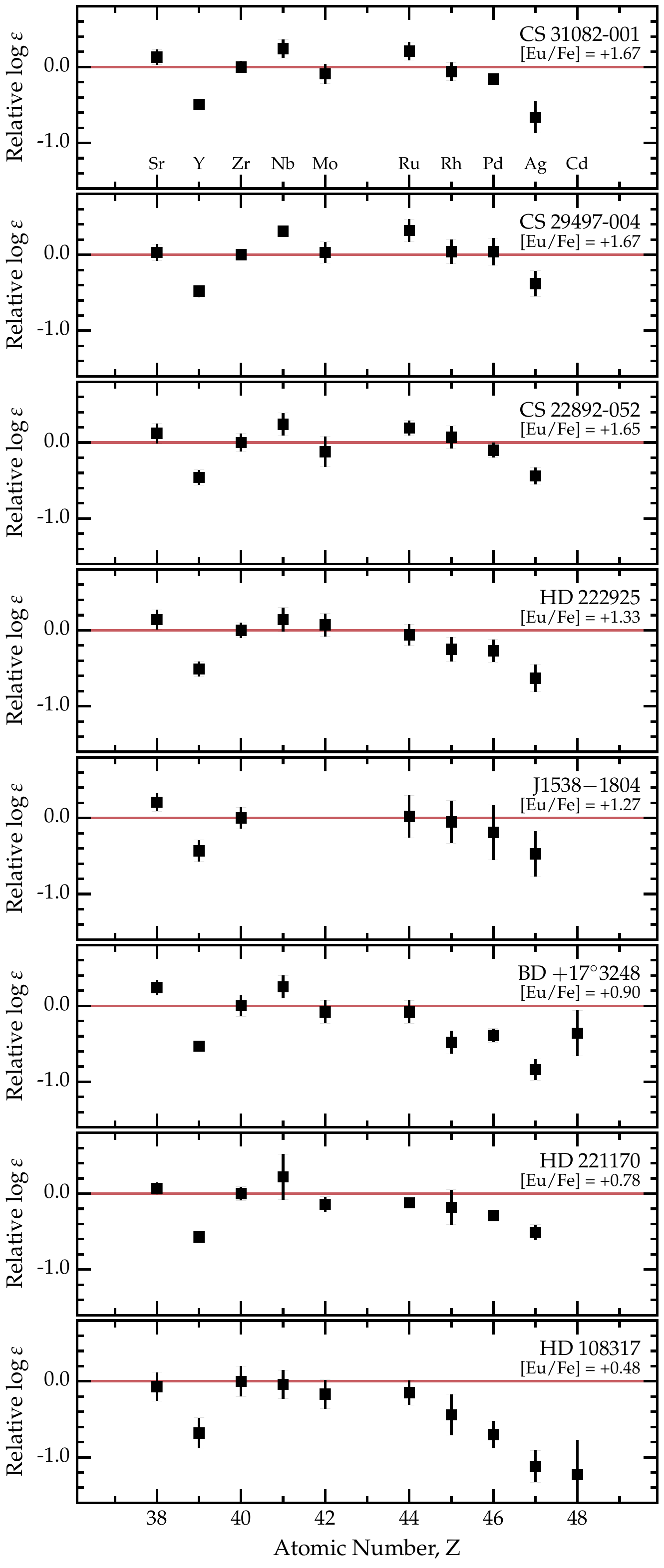}
\end{center}
\caption{
\label{sragplot}
Differences between the 38~$\leq Z \leq$~48 abundances
and the Solar \rpro\ residuals \citep{sneden08} 
for eight \rpro-enhanced stars.
The differences are normalized to Zr.
The red line in each panel indicates
perfect agreement with the \rpro\ residuals.
Data are taken from
\citet{siqueiramello13} for \object[BPS CS 31082-001]{CS~31082-001},
\citet{hill17} for \object[BPS CS 29497-004]{CS~29497-004},
\citet{sneden03a} for \object[BPS CS 22892-052]{CS~22892-052},
this study for \hdtwo,
\citet{sakari18a} for 
\object[2MASS J15383085-1804242]{J15383085$-$1804242},
\citet{cowan02} and \citet{roederer10b} for 
\object[BD +173248]{BD~$+$17\degree3248},
\citet{ivans06} for \object[HD 221170]{HD~221170}, and
\citet{roederer12d,roederer14d} for \object[HD 108317]{HD~108317}.
The stars are sorted by decreasing [Eu/Fe] ratios,
which are adopted from
\citet{sneden09} or the references given above.
 }
\end{figure}

The elements whose abundances
deviate significantly from the scaled \rpro\ pattern
in \hdtwo---Y, Rh, Pd, and Ag---deviate in a manner consistent with 
behavior observed in other \rpro-enhanced stars
(e.g., \citealt{johnson02rpro}).
Figure~\ref{sragplot} illustrates this point.
Eight \rpro-enhanced stars are illustrated in Figure~\ref{sragplot},
and they have been selected for inclusion because
Ru, Rh, Pd, and Ag (44~$\leq Z \leq$~47) have been detected.
Cd ($Z =$~48) has also been detected in two of them.
The points in Figure~\ref{sragplot} represent the
differences between the 
$\log\varepsilon$ abundances and the Solar \rpro\ residuals,
and these points have been normalized to Zr.

The Y abundances 
consistently deviate by
$\approx -$0.5~dex from the
scaled Solar \rpro\ residuals.
The downward trend relative to the
scaled \rpro\ residuals for 44~$\leq Z \leq$~47 
appears to be a distinct characteristic of the \rpro\ signature.
It is unclear whether this trend extends to Cd, 
because Cd has been detected in so few stars.
The deviations 
for the 44~$\leq Z \leq$~47~elements
also become more pronounced
as the [Eu/Fe] (or [Zr/Fe]) ratios decrease,
which reaffirms the conclusion of \citet{hansen12}
that production of these elements is not always fully
coupled to Zr or Eu.
The similar abundance ratios among the
38~$\leq Z \leq$~47~elements for the
stars in the top five panels of
Figure~\ref{sragplot} indicate 
relatively robust production ratios among the
\rpro\ events that enriched these \rtwo\ stars.

% KEEP THIS TEXT IN THE COMMENTS JUST IN CASE IT BECOMES USEFUL.
%
%These elements are slightly less massive than the
%$A \gtrsim$~110 mass region where
%fission fragments
%from trans-uranic nuclei are predicted to
%impact the final abundances
%(e.g., \citealt{goriely13,shibagaki16}).
%Empirically, only one of these eight stars
%shown in Figure~\ref{sragplot}
%(\object[BPS CS 31082-001]{CS~31082-001}) has a significant
%actinide boost \citet{hill02},
%so there is insufficient information available
%to comment on whether the observed effects
%correlate in any way with actinide production.
%Additional data for these elements in 
%\rpro-enhanced stars would be welcome.

\subsubsection{The Actinides}
\label{actinides}

The $^{232}$Th isotope is the only heavy, radioactive isotope
detected in \hdtwo.
Five lines of Th~\textsc{ii} are detected, and
all give concordant abundances.
The $^{238}$U isotope cannot be detected
because blends 
compromise U~\textsc{ii} lines that might otherwise
be detectable at 
3550.82~\AA\ (blended with La~\textsc{ii}),
3859.57~\AA\ (Fe~\textsc{i}),
4050.13~\AA\ (La~\textsc{ii}),
4090.13~\AA\ (Fe~\textsc{i}), and
4241.66~\AA\ (Zr~\textsc{i}).

The actinide elements Th and U
can only be produced by \rpro\ nucleosynthesis,
but their production is not well-understood theoretically at present.
There appears to be a genuine dispersion
among the $\log\varepsilon$(Th/Eu) ratios of metal-poor,
\rpro\ enhanced stars,
ranging from $-$0.12
(\object[2MASS J09544277+5246414]{2MASS~J0954$+$5246}; 
\citealt{holmbeck18rpro}) 
to $-$0.84
(\object[DES J033523.85-540407.5]{DES~J0335$-$5404}; 
\citealt{ji18th}).
The mean $\log\varepsilon$(Th/Eu) ratio in \hdtwo\ 
lies between these extremes, 
$-$0.44~$\pm$~0.14.
For completeness, we note that the
$\log\varepsilon$(Th/Eu) ratio 
derived from only the Th~\textsc{ii} line at 4019.13~\AA,
which is often the only Th abundance indicator
in other stars, is $-$0.61 in \hdtwo; this ratio
closely matches that found in other \rtwo\ stars.

The dispersion among the observed ratios
is known as the ``actinide boost''
\citep{hill02,schatz02}.
This phenomenon results in 
enhanced abundances of Th and U relative to
levels expected based on the low metallicities
(i.e., old ages)
and predictions for actinide production
relative to lighter, stable isotopes.
Recent observations of the brightest 
\rpro-enhanced star in \rettwolong\ by \citet{ji18th}
suggest the existence of actinide-deficient stars as well.
Attempts to characterize actinide production using
globular clusters of known ages have so far been 
unsuccessful due to the large observational uncertainties,
small number of clusters studied, and the 
lack of actinide-boost signatures in any 
cluster studied 
\citep{roederer15,roederer16a}.
New theoretical efforts to understand the physical
nature of the actinide boost phenomenon 
(e.g., \citealt{holmbeck18actinide})
are most welcome.
We thus refrain
from estimating the age of \hdtwo\ from
the radioactive decay of $^{232}$Th.

\subsection{The Environment of HD 222925}
\label{kinematics}

In this section we consider 
additional kinematic and chemical information
to infer possible origin scenarios for \hdtwo.
\citet{roederer18c} identified \hdtwo\ as a member of
a group of kinematically-similar \rpro-enhanced stars.
This group and others were identified 
using only the stars' 
specific orbital energy and action integrals.
Chemistry played no role in the group definitions,
yet the [Fe/H] dispersion of each group was
considerably smaller than would be expected
if the groups were selected at random among the
\rpro-enhanced stars considered.
The [Fe/H] dispersions of these groups are comparable to
or smaller than that found among the \rpro-enhanced stars
in the \rettwolong\ dwarf galaxy.
Other low-mass dwarf galaxies typically show
moderately small [Fe/H] dispersions
($\approx$~0.3--0.6~dex; e.g., \citealt{kirby11mdf}).
This line of reasoning led \citeauthor{roederer18c}\ 
to conclude that \rpro-enhanced stars within
each candidate group may share a common origin.

The two other stars that are candidates for membership in 
the group with \hdtwo,
\object[HD 20]{HD~20} 
and
\object[2MASS J01530024-3417360]{J0153$-$3417} $=$ \mbox{HD~11582},
have similarly high [Fe/H] ratios, $-$1.58 and $-$1.50
\citep{barklem05heres,hansen18}.
Their [Eu/H] ratios,  
$-$0.78 and $-$0.79,
are high, but lower than \hdtwo.
The stars in this group have highly eccentric, retrograde
orbits that pass within $\sim$~1~kpc
of the Galactic center and extend 
to $\sim$~15~kpc from the Galactic center.
Any stellar system with such a small pericentric radius
would be quickly tidally disrupted by the Milky Way, 
so chemical evolution in the progenitor would have been 
truncated as soon as it was accreted.
This conclusion is consistent with our finding that 
the metals in \hdtwo\ were produced on short
timescales by Type~II supernovae.

The metallicity dispersion of the three stars
is small and consistent with zero,
like that of most globular clusters.
It is unlikely, however, that the progenitor was a globular cluster.
Neither \hdtwo\ nor the other stars in this group
exhibit the light-element chemical signatures 
among O, Na, Mg, or Al that are commonly
found among the majority of globular cluster stars
\citep{barklem05heres,roederer14c}.
Furthermore, no star is known in any
globular cluster with [Eu/Fe] as enhanced as \hdtwo.

\citet{roederer18c} instead proposed that these three stars may 
represent a few of the remnants from a dwarf galaxy.
The dwarf galaxy luminosity-metallicity relation
\citep{kirby08metal,walker16}
predicts a satellite progenitor with 
a mass or luminosity comparable to that of the
\object[NAME Sculptor Dwarf Galaxy]{Sculptor} dwarf galaxy
($M_{*} \approx 2.3 \times 10^{6}$~\msun; \citealt{mcconnachie12}).
We regard this as unlikely 
because few known field \rpro-enhanced stars
have such high metallicities.
Instead, we speculate that these three stars
may have formed in a 
relatively dense clump of gas in close proximity
to an \rpro\ event in the putative progenitor satellite.

A few moderately metal-poor
\rtwo\ stars are known in the low-luminosity
classical dwarf galaxies
\object[NAME DRACO DSPH]{Draco} (Dra; \citealt{cohen09dra})
and
\object[NAME URSA MINOR DWARF GALAXY]{Ursa Minor} 
(UMi; \citealt{shetrone01}).
One of those stars, 
\object[COS 82]{UMi COS~82} \citep{aoki07th},
has a metallicity 
([Fe/H]~$= -$1.42) and 
Eu abundance
([Eu/H]~$= -$0.18) 
that are similar to \hdtwo.
There is a more continuous rise in [Eu/Fe] 
with increasing [Fe/H] in 
\object[NAME URSA MINOR DWARF GALAXY]{UMi} 
\citep{cohen10umi}, suggesting that 
multiple \rpro\ events may have occurred in this system.
Unlike \hdtwo, which has [$\alpha$/Fe]~$\approx +$0.4,
the [$\alpha$/Fe] ratios of 
\object[COS 82]{UMi COS~82} 
and other stars with [Fe/H]~$\approx -$1.5 in
\object[NAME DRACO DSPH]{Dra} 
and
\object[NAME URSA MINOR DWARF GALAXY]{UMi}
are sub-Solar or only slightly super-Solar \citep{sadakane04,kirby11alpha}.
This signals the presence of 
Type~Ia supernova ejecta, indicating that the
chemical enrichment timescales in 
\object[NAME DRACO DSPH]{Dra} 
and
\object[NAME URSA MINOR DWARF GALAXY]{UMi} 
were longer than in the progenitor system of \hdtwo.
The
longer timescales may have permitted multiple \rpro\
events to have occurred in
\object[NAME DRACO DSPH]{Dra} 
and
\object[NAME URSA MINOR DWARF GALAXY]{UMi} (cf.\ \citealt{tsujimoto17}).
\hdtwo\ probably did not form in systems like these.

It is probable that the \rpro\ material observed
in \hdtwo\ was produced by a single \rpro\ event
(cf.\ \citealt{ji16nat,tsujimoto17}).
Following \citet{roederer18c},
we estimate the 
mass of stars formed with \hdtwo\
using the derived [Eu/H] ratios, 
adopting an \rpro\ mass yield,
and making reasonable assumptions about the
star-formation efficiency and metal loss from the progenitor system.
A neutron star merger, like that associated with 
the GW170817 event, could eject
$\sim$~0.005~\msun\ of \rpro\ material at and beyond the second
\rpro\ peak (see discussion in \citealt{cote18rpro}).
Incorporating this ejecta into
$\sim 10^{3}$ to $10^{4}$~\msun\ of stars
formed after the merger
would produce the observed [Eu/H] ratios,
assuming a star formation efficiency of $\sim$~1\%
and no loss of \rpro\ metals.
This mass is comparable to that of
the lowest-mass dwarf galaxies known today, 
including 
\rettwolong\ \citep{ji16ret2} and 
\object[NAME TUCANA III]{Tucana~III} \citep{hansen17tuc3},
which are known to contain \rpro-enhanced stars.
There is considerable scatter 
among the [Eu/H] ratios of individual stars
in these systems
($-$1.6~$\leq$~[Eu/H]~$\leq -$0.3),
and the average [Eu/H] ratio of stars in a system
may provide a more meaningful measure than that from any one star.
Our order-of-magnitude estimate is suggestive, but of course
identifying other stars with kinematic properties
similar to \hdtwo\ 
would help characterize the 
chemical evolution and nature of this putative progenitor system.

One star with relatively high metallicity
([Fe/H]~$= -$1.67) and \rpro\ enhancement
([Eu/Fe]~$= +$0.99; [Eu/H]~$= -$0.68) has been found
along a line of sight toward the Milky Way bulge
(\object[2MASS J18174532-3353235]{2MASS~J18174532$-$3353235}; 
\citealt{johnson13}).
This star is $\alpha$ enhanced, like \hdtwo, and the \rpro\ pattern
of both stars closely matches that found in \rpro-enhanced stars
with $-$3.5~$\leq$~[Fe/H]~$\leq -$2.0.
The existence of stars like 
\object[2MASS J18174532-3353235]{2MASS~J18174532$-$3353235}
and \hdtwo---which are relatively metal-rich,
$\alpha$ enhanced, and contain large
proportions of \rpro\ elements---supports the early onset of the \rpro\ 
from a single class of progenitors,
without needing to invoke 
mixing of multiple progenitor types.

\subsection{Are There Other Misclassified \textit{R}-Process Enhanced Stars?}
\label{cpstars}

We noted in Section~\ref{intro}
that \citet{houk75}
classified \hdtwo\ as being a chemically peculiar
``Sr Eu'' star.
Are there other \rpro-enhanced stars in this catalog or subsequent ones
(\citealt{houk78,houk82}; \citealt{houk88}; \citealt{houk99})
that have been overlooked?
Houk and colleagues classified Ap (``peculiar A'') stars 
by the relative strengths of several metal lines, including
Cr~\textsc{ii} ($\lambda$4111, $\lambda$4171), 
Sr~\textsc{ii} ($\lambda$4077, $\lambda$4216), and 
Eu~\textsc{ii} ($\lambda$4128--30).
\hdtwo\ was not designated with ``Cr,'' which 
indicates its metal lines are relatively weak.
These catalogs list 36~stars 
designated with ``Eu'' but not ``Cr.''
Eleven of these stars have been identified as
$\alpha^{2}$~CVn variables
\citep{bernhard15}, which
have strong surface magnetic fields 
that stratify metals in the atmosphere.
One is a confirmed $\delta$~Scu variable
\citep{martinez02}.
Another 22 of these stars have \teff\ estimates
(\citealt{mcdonald12}; I.U.R., unpublished)
between $\sim$~6500 and 9500~K that
suggest they are not
late-type metal-poor stars.
Several of these stars are also well-studied
chemically-peculiar standards,
often exhibiting Zeeman splitting of metal absorption lines
(e.g.\ \citealt{freyhammer08}).
The only star among the 36 that is not described by any of these
characteristics is \hdtwo.
We conclude that it is unlikely that other
stars like \hdtwo\ have been
overlooked based on their
initial classification in the catalogs by Houk and colleagues.

\section{Conclusions}
\label{conclusions}

We present a detailed analysis of the stellar parameters
and abundances of the bright, metal-poor star \hdtwo.
This star is a field equivalent of the 
He core-burning 
RHB stars found in metal-poor
globular clusters.
\hdtwo\ is the brightest \rtwo\ star known ($V =$~9.02),
one of the warmest \rtwo\ stars known (\teff~$=$~5636~K),
the most metal-rich \rtwo\ star known ([Fe/H]~$= -$1.47),
and contains the highest abundance
of \rpro\ elements ([Eu/H]~$= -$0.14)
among \rpro-enhanced stars.

\hdtwo\ shows no evidence of binarity, and it is 
not enhanced in carbon.
\hdtwo\ is $\alpha$-enhanced,
with [(O, Mg, Si, Ca, Ti)/Fe]~$\approx +$0.4,
and the abundance ratios among elements with $Z \leq$~30
are consistent with other metal-poor field stars.
The heavy elements are highly enhanced in \hdtwo,
perhaps best exemplified by [Eu/Fe]~$= +$1.33.
The abundances of elements with $Z \geq$~38
closely match the scaled Solar \rpro\ residuals,
with no evidence for \spro\ contamination.
Deviations from the Solar \rpro\ pattern for
Y, Ru, Rh, Pd, and Ag ($Z =$~39, 44--47)
match those found in other \rpro-enhanced stars.
\hdtwo\ does not exhibit a strong actinide boost,
but the $\log\varepsilon$(Th/Eu) ratio, $-$0.44, is 
intermediate between stars with and without 
the actinide boost.

\hdtwo\ is a member of a group of \rpro-enhanced
stars with similar kinematics \citep{roederer18c}.
If we assume that the \rpro\ material in \hdtwo\ was
produced by a single, high-yield nucleosynthesis event,
like a neutron star merger,
we conclude that the progenitor system had
a stellar mass $\sim 10^{3}$--$10^{4}$,
comparable to the surviving population of 
ultra-faint dwarf galaxies.
This conclusion is consistent with our assertion that
the metals with $Z \leq$~30 in \hdtwo\ were produced
by Type~II supernovae with minimal contributions
from Type~Ia supernovae or AGB stars,
as is typical for the surviving population of ultra-faint dwarf galaxies
(e.g., \citealt{frebel14seg1,ji16tuc2}).
The existence of relatively metal-rich stars,
such as \hdtwo,
with \rpro\ abundance signatures that are excellent matches to \rtwo\
stars with $-$3.5~$\leq$~[Fe/H]~$\leq -$2.0,
supports the early onset of the \rpro\ 
from a single class of progenitors,
without needing to invoke 
mixing of multiple progenitor types.

\acknowledgments

We thank the referee for a prompt and constructive report.
I.U.R., V.M.P., T.C.B., R.E., and A.F.\
acknowledge financial support from
grant PHY~14-30152 (Physics Frontier Center/JINA-CEE)
awarded by the U.S.\ National Science Foundation (NSF).~
I.U.R.\ acknowledges support from NSF grants AST-1613536 and
AST-1815403.
C.M.S.\ acknowledges funding from the 
Kenilworth Fund of the New York Community Trust.
A.F.\ acknowledges support from NSF grant AST-1716251.
This research has made use of NASA's
Astrophysics Data System Bibliographic Services;
the arXiv pre-print server operated by Cornell University;
the SIMBAD and VizieR
databases hosted by the
Strasbourg Astronomical Data Center;
the ASD hosted by NIST;
%the National Institute of Standards and Technology;
%the MAST at STScI; 
and
Image Reduction and Analysis Facility (IRAF) software packages
distributed by the National Optical Astronomy Observatories,
which are operated by AURA,
under cooperative agreement with the NSF.~
This work has also made use of data from the European Space Agency (ESA)
mission {\it Gaia} 
(\url{http://www.cosmos.esa.int/gaia}), 
processed by the {\it Gaia} Data Processing and Analysis Consortium (DPAC,
\url{http://www.cosmos.esa.int/web/gaia/dpac/consortium}). 
Funding for the DPAC has been provided by national institutions, in particular
the institutions participating in the {\it Gaia} Multilateral Agreement.

\facility{Magellan (MIKE)}

\software{IRAF \citep{tody93},
matplotlib \citep{hunter07},
MOOG \citep{sneden73},
numpy \citep{vanderwalt11},
R \citep{rsoftware},
scipy \citep{jones01}}

\bibliographystyle{aasjournal}
\bibliography{iuroederer}

\begin{thebibliography}{}
\expandafter\ifx\csname natexlab\endcsname\relax\def\natexlab#1{#1}\fi
\providecommand{\url}[1]{\href{#1}{#1}}
\providecommand{\dodoi}[1]{doi:~\href{http://doi.org/#1}{\nolinkurl{#1}}}
\providecommand{\doeprint}[1]{\href{http://ascl.net/#1}{\nolinkurl{http://ascl.net/#1}}}
\providecommand{\doarXiv}[1]{\href{https://arxiv.org/abs/#1}{\nolinkurl{https://arxiv.org/abs/#1}}}

\bibitem[{{Abbott} {et~al.}(2017{\natexlab{a}}){Abbott}, {Abbott}, {Abbott},
  {Acernese}, {Ackley}, {Adams}, {Adams}, {Addesso}, {Adhikari}, {Adya}, \&
  et~al.}]{abbott17prl}
{Abbott}, B.~P., {Abbott}, R., {Abbott}, T.~D., {et~al.} 2017{\natexlab{a}},
  Physical Review Letters, 119, 161101, \dodoi{10.1103/PhysRevLett.119.161101}

\bibitem[{{Abbott} {et~al.}(2017{\natexlab{b}}){Abbott}, {Abbott}, {Abbott},
  {Acernese}, {Ackley}, {Adams}, {Adams}, {Addesso}, {Adhikari}, {Adya}, \&
  et~al.}]{abbott17multimessenger}
---. 2017{\natexlab{b}}, \apjl, 848, L12, \dodoi{10.3847/2041-8213/aa91c9}

\bibitem[{{Af{\c s}ar} {et~al.}(2018){Af{\c s}ar}, {Bozkurt}, {B{\"o}cek
  Topcu}, {Casetti-Dinescu}, {Sneden}, \& {{\c S}ehitog{\#773}lu}}]{afsar18}
{Af{\c s}ar}, M., {Bozkurt}, Z., {B{\"o}cek Topcu}, G., {et~al.} 2018, \aj,
  155, 240, \dodoi{10.3847/1538-3881/aabe86}

\bibitem[{{Af{\c s}ar} {et~al.}(2012){Af{\c s}ar}, {Sneden}, \&
  {For}}]{afsar12}
{Af{\c s}ar}, M., {Sneden}, C., \& {For}, B.-Q. 2012, \aj, 144, 20,
  \dodoi{10.1088/0004-6256/144/1/20}

\bibitem[{{Aldenius} {et~al.}(2009){Aldenius}, {Lundberg}, \&
  {Blackwell-Whitehead}}]{aldenius09}
{Aldenius}, M., {Lundberg}, H., \& {Blackwell-Whitehead}, R. 2009, \aap, 502,
  989, \dodoi{10.1051/0004-6361/200911844}

\bibitem[{{Alonso} {et~al.}(1999){Alonso}, {Arribas}, \&
  {Mart{\'{\i}}nez-Roger}}]{alonso99b}
{Alonso}, A., {Arribas}, S., \& {Mart{\'{\i}}nez-Roger}, C. 1999, \aaps, 140,
  261, \dodoi{10.1051/aas:1999521}

\bibitem[{{Andrievsky} {et~al.}(2007){Andrievsky}, {Spite}, {Korotin}, {Spite},
  {Bonifacio}, {Cayrel}, {Hill}, \& {Fran{\c c}ois}}]{andrievsky07}
{Andrievsky}, S.~M., {Spite}, M., {Korotin}, S.~A., {et~al.} 2007, \aap, 464,
  1081, \dodoi{10.1051/0004-6361:20066232}

\bibitem[{{Andrievsky} {et~al.}(2008){Andrievsky}, {Spite}, {Korotin}, {Spite},
  {Bonifacio}, {Cayrel}, {Hill}, \& {Fran{\c c}ois}}]{andrievsky08}
---. 2008, \aap, 481, 481, \dodoi{10.1051/0004-6361:20078837}

\bibitem[{{Aoki} {et~al.}(2007{\natexlab{a}}){Aoki}, {Beers}, {Christlieb},
  {Norris}, {Ryan}, \& {Tsangarides}}]{aoki07cemp}
{Aoki}, W., {Beers}, T.~C., {Christlieb}, N., {et~al.} 2007{\natexlab{a}},
  \apj, 655, 492, \dodoi{10.1086/509817}

\bibitem[{{Aoki} {et~al.}(2007{\natexlab{b}}){Aoki}, {Honda}, {Sadakane}, \&
  {Arimoto}}]{aoki07th}
{Aoki}, W., {Honda}, S., {Sadakane}, K., \& {Arimoto}, N. 2007{\natexlab{b}},
  \pasj, 59, L15, \dodoi{10.1093/pasj/59.3.L15}

\bibitem[{{Aoki} {et~al.}(2002){Aoki}, {Ryan}, {Norris}, {Beers}, {Ando}, \&
  {Tsangarides}}]{aoki02pb}
{Aoki}, W., {Ryan}, S.~G., {Norris}, J.~E., {et~al.} 2002, \apj, 580, 1149,
  \dodoi{10.1086/343885}

\bibitem[{{Asplund} {et~al.}(2009){Asplund}, {Grevesse}, {Sauval}, \&
  {Scott}}]{asplund09}
{Asplund}, M., {Grevesse}, N., {Sauval}, A.~J., \& {Scott}, P. 2009, \araa, 47,
  481, \dodoi{10.1146/annurev.astro.46.060407.145222}

\bibitem[{{Bailer-Jones} {et~al.}(2018){Bailer-Jones}, {Rybizki}, {Fouesneau},
  {Mantelet}, \& {Andrae}}]{bailerjones18}
{Bailer-Jones}, C.~A.~L., {Rybizki}, J., {Fouesneau}, M., {Mantelet}, G., \&
  {Andrae}, R. 2018, \aj, 156, 58, \dodoi{10.3847/1538-3881/aacb21}

\bibitem[{{Barklem} \& {Aspelund-Johansson}(2005)}]{barklem05feii}
{Barklem}, P.~S., \& {Aspelund-Johansson}, J. 2005, \aap, 435, 373,
  \dodoi{10.1051/0004-6361:20042469}

\bibitem[{{Barklem} {et~al.}(2000){Barklem}, {Piskunov}, \&
  {O'Mara}}]{barklem00h}
{Barklem}, P.~S., {Piskunov}, N., \& {O'Mara}, B.~J. 2000, \aap, 355, L5

\bibitem[{{Barklem} {et~al.}(2005){Barklem}, {Christlieb}, {Beers}, {Hill},
  {Bessell}, {Holmberg}, {Marsteller}, {Rossi}, {Zickgraf}, \&
  {Reimers}}]{barklem05heres}
{Barklem}, P.~S., {Christlieb}, N., {Beers}, T.~C., {et~al.} 2005, \aap, 439,
  129, \dodoi{10.1051/0004-6361:20052967}

\bibitem[{{Battistini} \& {Bensby}(2015)}]{battistini15}
{Battistini}, C., \& {Bensby}, T. 2015, \aap, 577, A9,
  \dodoi{10.1051/0004-6361/201425327}

\bibitem[{{Battistini} \& {Bensby}(2016)}]{battistini16}
---. 2016, \aap, 586, A49, \dodoi{10.1051/0004-6361/201527385}

\bibitem[{{Beers} \& {Christlieb}(2005)}]{beers05}
{Beers}, T.~C., \& {Christlieb}, N. 2005, \araa, 43, 531,
  \dodoi{10.1146/annurev.astro.42.053102.134057}

\bibitem[{{Beers} {et~al.}(2014){Beers}, {Norris}, {Placco}, {Lee}, {Rossi},
  {Carollo}, \& {Masseron}}]{beers14}
{Beers}, T.~C., {Norris}, J.~E., {Placco}, V.~M., {et~al.} 2014, \apj, 794, 58,
  \dodoi{10.1088/0004-637X/794/1/58}

\bibitem[{{Beers} {et~al.}(1992){Beers}, {Preston}, \& {Shectman}}]{beers92}
{Beers}, T.~C., {Preston}, G.~W., \& {Shectman}, S.~A. 1992, \aj, 103, 1987,
  \dodoi{10.1086/116207}

\bibitem[{{Belmonte} {et~al.}(2017){Belmonte}, {Pickering}, {Ruffoni}, {Den
  Hartog}, {Lawler}, {Guzman}, \& {Heiter}}]{belmonte17}
{Belmonte}, M.~T., {Pickering}, J.~C., {Ruffoni}, M.~P., {et~al.} 2017, \apj,
  848, 125, \dodoi{10.3847/1538-4357/aa8cd3}

\bibitem[{{Beniamini} {et~al.}(2016){Beniamini}, {Hotokezaka}, \&
  {Piran}}]{beniamini16b}
{Beniamini}, P., {Hotokezaka}, K., \& {Piran}, T. 2016, \apj, 832, 149,
  \dodoi{10.3847/0004-637X/832/2/149}

\bibitem[{{Bensby} {et~al.}(2014){Bensby}, {Feltzing}, \& {Oey}}]{bensby14}
{Bensby}, T., {Feltzing}, S., \& {Oey}, M.~S. 2014, \aap, 562, A71,
  \dodoi{10.1051/0004-6361/201322631}

\bibitem[{{Bergemann}(2011)}]{bergemann11}
{Bergemann}, M. 2011, \mnras, 413, 2184,
  \dodoi{10.1111/j.1365-2966.2011.18295.x}

\bibitem[{{Bergemann} \& {Gehren}(2008)}]{bergemann08}
{Bergemann}, M., \& {Gehren}, T. 2008, \aap, 492, 823,
  \dodoi{10.1051/0004-6361:200810098}

\bibitem[{{Bergemann} {et~al.}(2012){Bergemann}, {Lind}, {Collet}, {Magic}, \&
  {Asplund}}]{bergemann12}
{Bergemann}, M., {Lind}, K., {Collet}, R., {Magic}, Z., \& {Asplund}, M. 2012,
  \mnras, 427, 27, \dodoi{10.1111/j.1365-2966.2012.21687.x}

\bibitem[{{Bernhard} {et~al.}(2015){Bernhard}, {H{\"u}mmerich}, {Otero}, \&
  {Paunzen}}]{bernhard15}
{Bernhard}, K., {H{\"u}mmerich}, S., {Otero}, S., \& {Paunzen}, E. 2015, \aap,
  581, A138, \dodoi{10.1051/0004-6361/201526424}

\bibitem[{{Bernstein} {et~al.}(2003){Bernstein}, {Shectman}, {Gunnels},
  {Mochnacki}, \& {Athey}}]{bernstein03}
{Bernstein}, R., {Shectman}, S.~A., {Gunnels}, S.~M., {Mochnacki}, S., \&
  {Athey}, A.~E. 2003, in \procspie, Vol. 4841, Instrument Design and
  Performance for Optical/Infrared Ground-based Telescopes, ed. M.~{Iye} \&
  A.~F.~M. {Moorwood}, 1694--1704

\bibitem[{{Bi{\'e}mont} {et~al.}(2000){Bi{\'e}mont}, {Garnir}, {Palmeri}, {Li},
  \& {Svanberg}}]{biemont00}
{Bi{\'e}mont}, E., {Garnir}, H.~P., {Palmeri}, P., {Li}, Z.~S., \& {Svanberg},
  S. 2000, \mnras, 312, 116, \dodoi{10.1046/j.1365-8711.2000.03094.x}

\bibitem[{{Bi{\'e}mont} {et~al.}(2011){Bi{\'e}mont}, {Blagoev}, {Engstr{\"o}m},
  {Hartman}, {Lundberg}, {Malcheva}, {Nilsson}, {Whitehead}, {Palmeri}, \&
  {Quinet}}]{biemont11}
{Bi{\'e}mont}, {\'E}., {Blagoev}, K., {Engstr{\"o}m}, L., {et~al.} 2011,
  \mnras, 414, 3350, \dodoi{10.1111/j.1365-2966.2011.18637.x}

\bibitem[{{Cain} {et~al.}(2018){Cain}, {Frebel}, {Gull}, {Ji}, {Placco},
  {Beers}, {Melendez}, {Ezzeddine}, {Casey}, {Hansen}, {Roederer}, \&
  {Sakari}}]{cain18}
{Cain}, M., {Frebel}, A., {Gull}, M., {et~al.} 2018, ArXiv e-prints.
\newblock \doarXiv{1807.03734}

\bibitem[{{Casagrande} {et~al.}(2010){Casagrande}, {Ram{\'{\i}}rez},
  {Mel{\'e}ndez}, {Bessell}, \& {Asplund}}]{casagrande10}
{Casagrande}, L., {Ram{\'{\i}}rez}, I., {Mel{\'e}ndez}, J., {Bessell}, M., \&
  {Asplund}, M. 2010, \aap, 512, A54, \dodoi{10.1051/0004-6361/200913204}

\bibitem[{{Casagrande} \& {VandenBerg}(2014)}]{casagrande14c}
{Casagrande}, L., \& {VandenBerg}, D.~A. 2014, \mnras, 444, 392,
  \dodoi{10.1093/mnras/stu1476}

\bibitem[{{Casagrande} {et~al.}(2014){Casagrande}, {Portinari}, {Glass},
  {Laney}, {Silva Aguirre}, {Datson}, {Andersen}, {Nordstr{\"o}m}, {Holmberg},
  {Flynn}, \& {Asplund}}]{casagrande14b}
{Casagrande}, L., {Portinari}, L., {Glass}, I.~S., {et~al.} 2014, \mnras, 439,
  2060, \dodoi{10.1093/mnras/stu089}

\bibitem[{{Castelli} \& {Kurucz}(2004)}]{castelli04}
{Castelli}, F., \& {Kurucz}, R.~L. 2004, ArXiv e-prints.
\newblock \doarXiv{astro-ph/0405087}

\bibitem[{{Cayrel}(1988)}]{cayrel88}
{Cayrel}, R. 1988, in IAU Symposium, Vol. 132, The Impact of Very High S/N
  Spectroscopy on Stellar Physics, ed. G.~{Cayrel de Strobel} \& M.~{Spite},
  345

\bibitem[{{Cayrel} {et~al.}(2004){Cayrel}, {Depagne}, {Spite}, {Hill}, {Spite},
  {Fran{\c c}ois}, {Plez}, {Beers}, {Primas}, {Andersen}, {Barbuy},
  {Bonifacio}, {Molaro}, \& {Nordstr{\"o}m}}]{cayrel04}
{Cayrel}, R., {Depagne}, E., {Spite}, M., {et~al.} 2004, \aap, 416, 1117,
  \dodoi{10.1051/0004-6361:20034074}

\bibitem[{{Cohen} {et~al.}(2008){Cohen}, {Christlieb}, {McWilliam}, {Shectman},
  {Thompson}, {Melendez}, {Wisotzki}, \& {Reimers}}]{cohen08emp}
{Cohen}, J.~G., {Christlieb}, N., {McWilliam}, A., {et~al.} 2008, \apj, 672,
  320, \dodoi{10.1086/523638}

\bibitem[{{Cohen} \& {Huang}(2009)}]{cohen09dra}
{Cohen}, J.~G., \& {Huang}, W. 2009, \apj, 701, 1053,
  \dodoi{10.1088/0004-637X/701/2/1053}

\bibitem[{{Cohen} \& {Huang}(2010)}]{cohen10umi}
---. 2010, \apj, 719, 931, \dodoi{10.1088/0004-637X/719/1/931}

\bibitem[{{C{\^o}t{\'e}} {et~al.}(2018){C{\^o}t{\'e}}, {Fryer}, {Belczynski},
  {Korobkin}, {Chru{\'s}li{\'n}ska}, {Vassh}, {Mumpower}, {Lippuner},
  {Sprouse}, {Surman}, \& {Wollaeger}}]{cote18rpro}
{C{\^o}t{\'e}}, B., {Fryer}, C.~L., {Belczynski}, K., {et~al.} 2018, \apj, 855,
  99, \dodoi{10.3847/1538-4357/aaad67}

\bibitem[{{Cowan} {et~al.}(2002){Cowan}, {Sneden}, {Burles}, {Ivans}, {Beers},
  {Truran}, {Lawler}, {Primas}, {Fuller}, {Pfeiffer}, \& {Kratz}}]{cowan02}
{Cowan}, J.~J., {Sneden}, C., {Burles}, S., {et~al.} 2002, \apj, 572, 861,
  \dodoi{10.1086/340347}

\bibitem[{{Cowan} {et~al.}(2005){Cowan}, {Sneden}, {Beers}, {Lawler},
  {Simmerer}, {Truran}, {Primas}, {Collier}, \& {Burles}}]{cowan05}
{Cowan}, J.~J., {Sneden}, C., {Beers}, T.~C., {et~al.} 2005, \apj, 627, 238,
  \dodoi{10.1086/429952}

\bibitem[{{Cowperthwaite} {et~al.}(2017){Cowperthwaite}, {Berger}, {Villar},
  {Metzger}, {Nicholl}, {Chornock}, {Blanchard}, {Fong}, {Margutti},
  {Soares-Santos}, {Alexander}, {Allam}, {Annis}, {Brout}, {Brown}, {Butler},
  {Chen}, {Diehl}, {Doctor}, {Drout}, {Eftekhari}, {Farr}, {Finley}, {Foley},
  {Frieman}, {Fryer}, {Garc{\'{\i}}a-Bellido}, {Gill}, {Guillochon}, {Herner},
  {Holz}, {Kasen}, {Kessler}, {Marriner}, {Matheson}, {Neilsen}, {Quataert},
  {Palmese}, {Rest}, {Sako}, {Scolnic}, {Smith}, {Tucker}, {Williams},
  {Balbinot}, {Carlin}, {Cook}, {Durret}, {Li}, {Lopes}, {Louren{\c c}o},
  {Marshall}, {Medina}, {Muir}, {Mu{\~n}oz}, {Sauseda}, {Schlegel}, {Secco},
  {Vivas}, {Wester}, {Zenteno}, {Zhang}, {Abbott}, {Banerji}, {Bechtol},
  {Benoit-L{\'e}vy}, {Bertin}, {Buckley-Geer}, {Burke}, {Capozzi}, {Carnero
  Rosell}, {Carrasco Kind}, {Castander}, {Crocce}, {Cunha}, {D'Andrea}, {da
  Costa}, {Davis}, {DePoy}, {Desai}, {Dietrich}, {Drlica-Wagner}, {Eifler},
  {Evrard}, {Fernandez}, {Flaugher}, {Fosalba}, {Gaztanaga}, {Gerdes},
  {Giannantonio}, {Goldstein}, {Gruen}, {Gruendl}, {Gutierrez}, {Honscheid},
  {Jain}, {James}, {Jeltema}, {Johnson}, {Johnson}, {Kent}, {Krause}, {Kron},
  {Kuehn}, {Nuropatkin}, {Lahav}, {Lima}, {Lin}, {Maia}, {March}, {Martini},
  {McMahon}, {Menanteau}, {Miller}, {Miquel}, {Mohr}, {Neilsen}, {Nichol},
  {Ogando}, {Plazas}, {Roe}, {Romer}, {Roodman}, {Rykoff}, {Sanchez},
  {Scarpine}, {Schindler}, {Schubnell}, {Sevilla-Noarbe}, {Smith}, {Smith},
  {Sobreira}, {Suchyta}, {Swanson}, {Tarle}, {Thomas}, {Thomas}, {Troxel},
  {Vikram}, {Walker}, {Wechsler}, {Weller}, {Yanny}, \&
  {Zuntz}}]{cowperthwaite17}
{Cowperthwaite}, P.~S., {Berger}, E., {Villar}, V.~A., {et~al.} 2017, \apjl,
  848, L17, \dodoi{10.3847/2041-8213/aa8fc7}

\bibitem[{{Cutri} {et~al.}(2003){Cutri}, {Skrutskie}, {van Dyk}, {Beichman},
  {Carpenter}, {Chester}, {Cambresy}, {Evans}, {Fowler}, {Gizis}, {Howard},
  {Huchra}, {Jarrett}, {Kopan}, {Kirkpatrick}, {Light}, {Marsh}, {McCallon},
  {Schneider}, {Stiening}, {Sykes}, {Weinberg}, {Wheaton}, {Wheelock}, \&
  {Zacarias}}]{cutri03}
{Cutri}, R.~M., {Skrutskie}, M.~F., {van Dyk}, S., {et~al.} 2003, VizieR Online
  Data Catalog, 2246

\bibitem[{{Den Hartog} {et~al.}(2003){Den Hartog}, {Lawler}, {Sneden}, \&
  {Cowan}}]{denhartog03}
{Den Hartog}, E.~A., {Lawler}, J.~E., {Sneden}, C., \& {Cowan}, J.~J. 2003,
  \apjs, 148, 543, \dodoi{10.1086/376940}

\bibitem[{{Den Hartog} {et~al.}(2006){Den Hartog}, {Lawler}, {Sneden}, \&
  {Cowan}}]{denhartog06}
---. 2006, \apjs, 167, 292, \dodoi{10.1086/508262}

\bibitem[{{Den Hartog} {et~al.}(2011){Den Hartog}, {Lawler}, {Sobeck},
  {Sneden}, \& {Cowan}}]{denhartog11}
{Den Hartog}, E.~A., {Lawler}, J.~E., {Sobeck}, J.~S., {Sneden}, C., \&
  {Cowan}, J.~J. 2011, \apjs, 194, 35, \dodoi{10.1088/0067-0049/194/2/35}

\bibitem[{{Den Hartog} {et~al.}(2014){Den Hartog}, {Ruffoni}, {Lawler},
  {Pickering}, {Lind}, \& {Brewer}}]{denhartog14}
{Den Hartog}, E.~A., {Ruffoni}, M.~P., {Lawler}, J.~E., {et~al.} 2014, \apjs,
  215, 23, \dodoi{10.1088/0067-0049/215/2/23}

\bibitem[{{Drout} {et~al.}(2017){Drout}, {Piro}, {Shappee}, {Kilpatrick},
  {Simon}, {Contreras}, {Coulter}, {Foley}, {Siebert}, {Morrell}, {Boutsia},
  {Di Mille}, {Holoien}, {Kasen}, {Kollmeier}, {Madore}, {Monson},
  {Murguia-Berthier}, {Pan}, {Prochaska}, {Ramirez-Ruiz}, {Rest}, {Adams},
  {Alatalo}, {Ba{\~n}ados}, {Baughman}, {Beers}, {Bernstein}, {Bitsakis},
  {Campillay}, {Hansen}, {Higgs}, {Ji}, {Maravelias}, {Marshall}, {Bidin},
  {Prieto}, {Rasmussen}, {Rojas-Bravo}, {Strom}, {Ulloa},
  {Vargas-Gonz{\'a}lez}, {Wan}, \& {Whitten}}]{drout17}
{Drout}, M.~R., {Piro}, A.~L., {Shappee}, B.~J., {et~al.} 2017, Science, 358,
  1570, \dodoi{10.1126/science.aaq0049}

\bibitem[{{Duquette} \& {Lawler}(1985)}]{duquette85}
{Duquette}, D.~W., \& {Lawler}, J.~E. 1985, Journal of the Optical Society of
  America B Optical Physics, 2, 1948, \dodoi{10.1364/JOSAB.2.001948}

\bibitem[{{Ezzeddine} {et~al.}(2017){Ezzeddine}, {Frebel}, \&
  {Plez}}]{ezzeddine17}
{Ezzeddine}, R., {Frebel}, A., \& {Plez}, B. 2017, \apj, 847, 142,
  \dodoi{10.3847/1538-4357/aa8875}

\bibitem[{{For} \& {Sneden}(2010)}]{for10}
{For}, B.-Q., \& {Sneden}, C. 2010, \aj, 140, 1694,
  \dodoi{10.1088/0004-6256/140/6/1694}

\bibitem[{{Frebel}(2018)}]{frebel18}
{Frebel}, A. 2018, ArXiv e-prints.
\newblock \doarXiv{1806.08955}

\bibitem[{{Frebel} {et~al.}(2008){Frebel}, {Collet}, {Eriksson}, {Christlieb},
  \& {Aoki}}]{frebel08he}
{Frebel}, A., {Collet}, R., {Eriksson}, K., {Christlieb}, N., \& {Aoki}, W.
  2008, \apj, 684, 588, \dodoi{10.1086/590327}

\bibitem[{{Frebel} {et~al.}(2014){Frebel}, {Simon}, \& {Kirby}}]{frebel14seg1}
{Frebel}, A., {Simon}, J.~D., \& {Kirby}, E.~N. 2014, \apj, 786, 74,
  \dodoi{10.1088/0004-637X/786/1/74}

\bibitem[{{Freyhammer} {et~al.}(2008){Freyhammer}, {Elkin}, {Kurtz}, {Mathys},
  \& {Martinez}}]{freyhammer08}
{Freyhammer}, L.~M., {Elkin}, V.~G., {Kurtz}, D.~W., {Mathys}, G., \&
  {Martinez}, P. 2008, \mnras, 389, 441,
  \dodoi{10.1111/j.1365-2966.2008.13595.x}

\bibitem[{{Gratton} {et~al.}(2000){Gratton}, {Sneden}, {Carretta}, \&
  {Bragaglia}}]{gratton00}
{Gratton}, R.~G., {Sneden}, C., {Carretta}, E., \& {Bragaglia}, A. 2000, \aap,
  354, 169

\bibitem[{{Gull} {et~al.}(2018){Gull}, {Frebel}, {Cain}, {Placco}, {Ji},
  {Abate}, {Ezzeddine}, {Karakas}, {Hansen}, {Sakari}, {Holmbeck}, {Santucci},
  {Casey}, \& {Beers}}]{gull18}
{Gull}, M., {Frebel}, A., {Cain}, M.~G., {et~al.} 2018, \apj, 862, 174,
  \dodoi{10.3847/1538-4357/aacbc3}

\bibitem[{{Hansen} {et~al.}(2013){Hansen}, {Bergemann}, {Cescutti}, {Fran{\c
  c}ois}, {Arcones}, {Karakas}, {Lind}, \& {Chiappini}}]{hansen13sr}
{Hansen}, C.~J., {Bergemann}, M., {Cescutti}, G., {et~al.} 2013, \aap, 551,
  A57, \dodoi{10.1051/0004-6361/201220584}

\bibitem[{{Hansen} {et~al.}(2012){Hansen}, {Primas}, {Hartman}, {Kratz},
  {Wanajo}, {Leibundgut}, {Farouqi}, {Hallmann}, {Christlieb}, \&
  {Nilsson}}]{hansen12}
{Hansen}, C.~J., {Primas}, F., {Hartman}, H., {et~al.} 2012, \aap, 545, A31,
  \dodoi{10.1051/0004-6361/201118643}

\bibitem[{{Hansen} {et~al.}(2017){Hansen}, {Simon}, {Marshall}, {Li},
  {Carollo}, {DePoy}, {Nagasawa}, {Bernstein}, {Drlica-Wagner}, {Abdalla},
  {Allam}, {Annis}, {Bechtol}, {Benoit-L{\'e}vy}, {Brooks}, {Buckley-Geer},
  {Carnero Rosell}, {Carrasco Kind}, {Carretero}, {Cunha}, {da Costa}, {Desai},
  {Eifler}, {Fausti Neto}, {Flaugher}, {Frieman}, {Garc{\'{\i}}a-Bellido},
  {Gaztanaga}, {Gerdes}, {Gruen}, {Gruendl}, {Gschwend}, {Gutierrez}, {James},
  {Krause}, {Kuehn}, {Kuropatkin}, {Lahav}, {Miquel}, {Plazas}, {Romer},
  {Sanchez}, {Santiago}, {Scarpine}, {Smith}, {Soares-Santos}, {Sobreira},
  {Suchyta}, {Swanson}, {Tarle}, {Walker}, \& {DES
  Collaboration}}]{hansen17tuc3}
{Hansen}, T.~T., {Simon}, J.~D., {Marshall}, J.~L., {et~al.} 2017, \apj, 838,
  44, \dodoi{10.3847/1538-4357/aa634a}

\bibitem[{{Hansen} {et~al.}(2018){Hansen}, {Holmbeck}, {Beers}, {Placco},
  {Roederer}, {Frebel}, {Sakari}, {Simon}, \& {Thompson}}]{hansen18}
{Hansen}, T.~T., {Holmbeck}, E.~M., {Beers}, T.~C., {et~al.} 2018, \apj, 858,
  92, \dodoi{10.3847/1538-4357/aabacc}

\bibitem[{{Hill} {et~al.}(2017){Hill}, {Christlieb}, {Beers}, {Barklem},
  {Kratz}, {Nordstr{\"o}m}, {Pfeiffer}, \& {Farouqi}}]{hill17}
{Hill}, V., {Christlieb}, N., {Beers}, T.~C., {et~al.} 2017, \aap, 607, A91,
  \dodoi{10.1051/0004-6361/201629092}

\bibitem[{{Hill} {et~al.}(2002){Hill}, {Plez}, {Cayrel}, {Beers},
  {Nordstr{\"o}m}, {Andersen}, {Spite}, {Spite}, {Barbuy}, {Bonifacio},
  {Depagne}, {Fran{\c c}ois}, \& {Primas}}]{hill02}
{Hill}, V., {Plez}, B., {Cayrel}, R., {et~al.} 2002, \aap, 387, 560,
  \dodoi{10.1051/0004-6361:20020434}

\bibitem[{{Holmbeck} {et~al.}(2018{\natexlab{a}}){Holmbeck}, {Surman},
  {Sprouse}, {Mumpower}, {Vassh}, {Beers}, \& {Kawano}}]{holmbeck18actinide}
{Holmbeck}, E.~M., {Surman}, R., {Sprouse}, T.~M., {et~al.} 2018{\natexlab{a}},
  ArXiv e-prints.
\newblock \doarXiv{1807.06662}

\bibitem[{{Holmbeck} {et~al.}(2018{\natexlab{b}}){Holmbeck}, {Beers},
  {Roederer}, {Placco}, {Hansen}, {Sakari}, {Sneden}, {Liu}, {Lee}, {Cowan}, \&
  {Frebel}}]{holmbeck18rpro}
{Holmbeck}, E.~M., {Beers}, T.~C., {Roederer}, I.~U., {et~al.}
  2018{\natexlab{b}}, \apjl, 859, L24, \dodoi{10.3847/2041-8213/aac722}

\bibitem[{{Horowitz} {et~al.}(2018){Horowitz}, {Arcones}, {C{\^o}t{\'e}},
  {Dillmann}, {Nazarewicz}, {Roederer}, {Schatz}, {Aprahamian}, {Atanasov},
  {Bauswein}, {Bliss}, {Brodeur}, {Clark}, {Frebel}, {Foucart}, {Hansen},
  {Just}, {Kankainen}, {McLaughlin}, {Kelly}, {Liddick}, {Lee}, {Lippuner},
  {Martin}, {Mendoza-Temis}, {Metzger}, {Mumpower}, {Perdikakis}, {Pereira},
  {O'Shea}, {Reifarth}, {Rogers}, {Siegel}, {Spyrou}, {Surman}, {Tang},
  {Uesaka}, \& {Wang}}]{horowitz18}
{Horowitz}, C.~J., {Arcones}, A., {C{\^o}t{\'e}}, B., {et~al.} 2018, ArXiv
  e-prints.
\newblock \doarXiv{1805.04637}

\bibitem[{{Houk}(1978)}]{houk78}
{Houk}, N. 1978, {Michigan catalogue of two-dimensional spectral types for the
  HD stars} (Univ. of Michigan)

\bibitem[{{Houk}(1982)}]{houk82}
---. 1982, {Michigan Catalogue of Two-dimensional Spectral Types for the HD
  stars. Volume 3. Declinations -40 to -26 Degrees.} (Univ. of Michigan)

\bibitem[{{Houk} \& {Cowley}(1975)}]{houk75}
{Houk}, N., \& {Cowley}, A.~P. 1975, {University of Michigan Catalogue of
  two-dimensional spectral types for the HD stars. Volume I. Declinations -90
  to -53 Degrees.} (Univ. of Michigan)

\bibitem[{{Houk} \& {Smith-Moore}(1988)}]{houk88}
{Houk}, N., \& {Smith-Moore}, M. 1988, {Michigan Catalogue of Two-dimensional
  Spectral Types for the HD Stars. Volume 4, Declinations -26 to -12 Degrees.}
  (Univ. of Michigan)

\bibitem[{{Houk} \& {Swift}(1999)}]{houk99}
{Houk}, N., \& {Swift}, C. 1999, {Michigan catalogue of two-dimensional
  spectral types for the HD Stars ; vol. 5} (Univ. of Michigan)

\bibitem[{{Hunter}(2007)}]{hunter07}
{Hunter}, J.~D. 2007, Computing in Science and Engineering, 9, 90,
  \dodoi{10.1109/MCSE.2007.55}

\bibitem[{{Ivans} {et~al.}(2006){Ivans}, {Simmerer}, {Sneden}, {Lawler},
  {Cowan}, {Gallino}, \& {Bisterzo}}]{ivans06}
{Ivans}, I.~I., {Simmerer}, J., {Sneden}, C., {et~al.} 2006, \apj, 645, 613,
  \dodoi{10.1086/504069}

\bibitem[{{Ivarsson} {et~al.}(2001){Ivarsson}, {Litz{\'e}n}, \&
  {Wahlgren}}]{ivarsson01}
{Ivarsson}, S., {Litz{\'e}n}, U., \& {Wahlgren}, G.~M. 2001, \physscr, 64, 455,
  \dodoi{10.1238/Physica.Regular.064a00455}

\bibitem[{{Jacobson} {et~al.}(2015){Jacobson}, {Keller}, {Frebel}, {Casey},
  {Asplund}, {Bessell}, {Da Costa}, {Lind}, {Marino}, {Norris}, {Pe{\~n}a},
  {Schmidt}, {Tisserand}, {Walsh}, {Yong}, \& {Yu}}]{jacobson15smss}
{Jacobson}, H.~R., {Keller}, S., {Frebel}, A., {et~al.} 2015, \apj, 807, 171,
  \dodoi{10.1088/0004-637X/807/2/171}

\bibitem[{{Ji} \& {Frebel}(2018)}]{ji18th}
{Ji}, A.~P., \& {Frebel}, A. 2018, \apj, 856, 138,
  \dodoi{10.3847/1538-4357/aab14a}

\bibitem[{{Ji} {et~al.}(2016{\natexlab{a}}){Ji}, {Frebel}, {Chiti}, \&
  {Simon}}]{ji16nat}
{Ji}, A.~P., {Frebel}, A., {Chiti}, A., \& {Simon}, J.~D. 2016{\natexlab{a}},
  \nat, 531, 610, \dodoi{10.1038/nature17425}

\bibitem[{{Ji} {et~al.}(2016{\natexlab{b}}){Ji}, {Frebel}, {Ezzeddine}, \&
  {Casey}}]{ji16tuc2}
{Ji}, A.~P., {Frebel}, A., {Ezzeddine}, R., \& {Casey}, A.~R.
  2016{\natexlab{b}}, \apjl, 832, L3, \dodoi{10.3847/2041-8205/832/1/L3}

\bibitem[{{Ji} {et~al.}(2016{\natexlab{c}}){Ji}, {Frebel}, {Simon}, \&
  {Chiti}}]{ji16ret2}
{Ji}, A.~P., {Frebel}, A., {Simon}, J.~D., \& {Chiti}, A. 2016{\natexlab{c}},
  \apj, 830, 93, \dodoi{10.3847/0004-637X/830/2/93}

\bibitem[{{Johnson} {et~al.}(2013){Johnson}, {McWilliam}, \&
  {Rich}}]{johnson13}
{Johnson}, C.~I., {McWilliam}, A., \& {Rich}, R.~M. 2013, \apjl, 775, L27,
  \dodoi{10.1088/2041-8205/775/1/L27}

\bibitem[{{Johnson} \& {Bolte}(2002)}]{johnson02rpro}
{Johnson}, J.~A., \& {Bolte}, M. 2002, \apj, 579, 616, \dodoi{10.1086/342829}

\bibitem[{{Jones} {et~al.}(2001){Jones}, {Oliphant}, \& {Peterson}}]{jones01}
{Jones}, E., {Oliphant}, T., \& {Peterson}, P., e.~a. 2001, {SciPy}: Open
  source scientific tools for {Python}.
\newblock \url{http://www.scipy.org/}

\bibitem[{{Kasen} {et~al.}(2017){Kasen}, {Metzger}, {Barnes}, {Quataert}, \&
  {Ramirez-Ruiz}}]{kasen17}
{Kasen}, D., {Metzger}, B., {Barnes}, J., {Quataert}, E., \& {Ramirez-Ruiz}, E.
  2017, \nat, 551, 80, \dodoi{10.1038/nature24453}

\bibitem[{{Katz} {et~al.}(2018){Katz}, {Sartoretti}, {Cropper}, {Panuzzo},
  {Seabroke}, {Viala}, {Benson}, {Blomme}, {Jasniewicz}, {Jean-Antoine},
  {Huckle}, {Smith}, {Baker}, {Crifo}, {Damerdji}, {David}, {Dolding},
  {Fr{\'e}mat}, {Gosset}, {Guerrier}, {Guy}, {Haigron}, {Jan{\ss}en},
  {Marchal}, {Plum}, {Soubiran}, {Th{\'e}venin}, {Ajaj}, {Allende Prieto},
  {Babusiaux}, {Boudreault}, {Chemin}, {Delle Luche}, {Fabre}, {Gueguen},
  {Hambly}, {Lasne}, {Meynadier}, {Pailler}, {Panem}, {Royer}, {Tauran},
  {Zurbach}, {Zwitter}, {Arenou}, {Bossini}, {Gomez}, {Lemaitre}, {Leclerc},
  {Morel}, {Munari}, {Turon}, {Vallenari}, \& {{\v Z}erjal}}]{katz18}
{Katz}, D., {Sartoretti}, P., {Cropper}, M., {et~al.} 2018, ArXiv e-prints.
\newblock \doarXiv{1804.09372}

\bibitem[{{Kelson}(2003)}]{kelson03}
{Kelson}, D.~D. 2003, \pasp, 115, 688, \dodoi{10.1086/375502}

\bibitem[{{Kelson} {et~al.}(2000){Kelson}, {Illingworth}, {van Dokkum}, \&
  {Franx}}]{kelson00}
{Kelson}, D.~D., {Illingworth}, G.~D., {van Dokkum}, P.~G., \& {Franx}, M.
  2000, \apj, 531, 159, \dodoi{10.1086/308445}

\bibitem[{{Kirby} {et~al.}(2011{\natexlab{a}}){Kirby}, {Cohen}, {Smith},
  {Majewski}, {Sohn}, \& {Guhathakurta}}]{kirby11alpha}
{Kirby}, E.~N., {Cohen}, J.~G., {Smith}, G.~H., {et~al.} 2011{\natexlab{a}},
  \apj, 727, 79, \dodoi{10.1088/0004-637X/727/2/79}

\bibitem[{{Kirby} {et~al.}(2011{\natexlab{b}}){Kirby}, {Lanfranchi}, {Simon},
  {Cohen}, \& {Guhathakurta}}]{kirby11mdf}
{Kirby}, E.~N., {Lanfranchi}, G.~A., {Simon}, J.~D., {Cohen}, J.~G., \&
  {Guhathakurta}, P. 2011{\natexlab{b}}, \apj, 727, 78,
  \dodoi{10.1088/0004-637X/727/2/78}

\bibitem[{{Kirby} {et~al.}(2008){Kirby}, {Simon}, {Geha}, {Guhathakurta}, \&
  {Frebel}}]{kirby08metal}
{Kirby}, E.~N., {Simon}, J.~D., {Geha}, M., {Guhathakurta}, P., \& {Frebel}, A.
  2008, \apjl, 685, L43, \dodoi{10.1086/592432}

\bibitem[{{Kordopatis} {et~al.}(2013){Kordopatis}, {Gilmore}, {Steinmetz},
  {Boeche}, {Seabroke}, {Siebert}, {Zwitter}, {Binney}, {de Laverny},
  {Recio-Blanco}, {Williams}, {Piffl}, {Enke}, {Roeser}, {Bijaoui}, {Wyse},
  {Freeman}, {Munari}, {Carrillo}, {Anguiano}, {Burton}, {Campbell}, {Cass},
  {Fiegert}, {Hartley}, {Parker}, {Reid}, {Ritter}, {Russell}, {Stupar},
  {Watson}, {Bienaym{\'e}}, {Bland-Hawthorn}, {Gerhard}, {Gibson}, {Grebel},
  {Helmi}, {Navarro}, {Conrad}, {Famaey}, {Faure}, {Just}, {Kos}, {Matijevi{\v
  c}}, {McMillan}, {Minchev}, {Scholz}, {Sharma}, {Siviero}, {de Boer}, \& {{\v
  Z}erjal}}]{kordopatis13rave}
{Kordopatis}, G., {Gilmore}, G., {Steinmetz}, M., {et~al.} 2013, \aj, 146, 134,
  \dodoi{10.1088/0004-6256/146/5/134}

\bibitem[{{Korotin} {et~al.}(2018){Korotin}, {Andrievsky}, \&
  {Zhukova}}]{korotin18}
{Korotin}, S.~A., {Andrievsky}, S.~M., \& {Zhukova}, A.~V. 2018, \mnras, 480,
  965, \dodoi{10.1093/mnras/sty1886}

\bibitem[{{Kramida} {et~al.}(2018){Kramida}, {Ralchenko}, {Reader}, \& {NIST
  ASD Team}}]{kramida18}
{Kramida}, A., {Ralchenko}, Y., {Reader}, J., \& {NIST ASD Team}. 2018, {NIST
  Atomic Spectra Database (ver. 5.5.6), [Online]. Available:
  {\tt{https://physics.nist.gov/asd}} National Institute of Standards and
  Technology, Gaithersburg, MD.}

\bibitem[{{Kunder} {et~al.}(2017){Kunder}, {Kordopatis}, {Steinmetz},
  {Zwitter}, {McMillan}, {Casagrande}, {Enke}, {Wojno}, {Valentini},
  {Chiappini}, {Matijevi{\v c}}, {Siviero}, {de Laverny}, {Recio-Blanco},
  {Bijaoui}, {Wyse}, {Binney}, {Grebel}, {Helmi}, {Jofre}, {Antoja}, {Gilmore},
  {Siebert}, {Famaey}, {Bienaym{\'e}}, {Gibson}, {Freeman}, {Navarro},
  {Munari}, {Seabroke}, {Anguiano}, {{\v Z}erjal}, {Minchev}, {Reid},
  {Bland-Hawthorn}, {Kos}, {Sharma}, {Watson}, {Parker}, {Scholz}, {Burton},
  {Cass}, {Hartley}, {Fiegert}, {Stupar}, {Ritter}, {Hawkins}, {Gerhard},
  {Chaplin}, {Davies}, {Elsworth}, {Lund}, {Miglio}, \& {Mosser}}]{kunder17}
{Kunder}, A., {Kordopatis}, G., {Steinmetz}, M., {et~al.} 2017, \aj, 153, 75,
  \dodoi{10.3847/1538-3881/153/2/75}

\bibitem[{{Kurucz} \& {Bell}(1995)}]{kurucz95}
{Kurucz}, R.~L., \& {Bell}, B. 1995, {Atomic line list} (Cambridge, MA:
  Smithsonian Astrophysical Observatory)

\bibitem[{{Lawler} {et~al.}(2001{\natexlab{a}}){Lawler}, {Bonvallet}, \&
  {Sneden}}]{lawler01la}
{Lawler}, J.~E., {Bonvallet}, G., \& {Sneden}, C. 2001{\natexlab{a}}, \apj,
  556, 452, \dodoi{10.1086/321549}

\bibitem[{{Lawler} \& {Dakin}(1989)}]{lawler89}
{Lawler}, J.~E., \& {Dakin}, J.~T. 1989, Journal of the Optical Society of
  America B Optical Physics, 6, 1457, \dodoi{10.1364/JOSAB.6.001457}

\bibitem[{{Lawler} {et~al.}(2007){Lawler}, {den Hartog}, {Labby}, {Sneden},
  {Cowan}, \& {Ivans}}]{lawler07}
{Lawler}, J.~E., {den Hartog}, E.~A., {Labby}, Z.~E., {et~al.} 2007, \apjs,
  169, 120, \dodoi{10.1086/510368}

\bibitem[{{Lawler} {et~al.}(2006){Lawler}, {Den Hartog}, {Sneden}, \&
  {Cowan}}]{lawler06}
{Lawler}, J.~E., {Den Hartog}, E.~A., {Sneden}, C., \& {Cowan}, J.~J. 2006,
  \apjs, 162, 227, \dodoi{10.1086/498213}

\bibitem[{{Lawler} {et~al.}(2018){Lawler}, {Feigenson}, {Sneden}, {Cowan}, \&
  {Nave}}]{lawler18}
{Lawler}, J.~E., {Feigenson}, T., {Sneden}, C., {Cowan}, J.~J., \& {Nave}, G.
  2018, ArXiv e-prints.
\newblock \doarXiv{1806.00581}

\bibitem[{{Lawler} {et~al.}(2013){Lawler}, {Guzman}, {Wood}, {Sneden}, \&
  {Cowan}}]{lawler13}
{Lawler}, J.~E., {Guzman}, A., {Wood}, M.~P., {Sneden}, C., \& {Cowan}, J.~J.
  2013, \apjs, 205, 11, \dodoi{10.1088/0067-0049/205/2/11}

\bibitem[{{Lawler} {et~al.}(2004){Lawler}, {Sneden}, \& {Cowan}}]{lawler04}
{Lawler}, J.~E., {Sneden}, C., \& {Cowan}, J.~J. 2004, \apj, 604, 850,
  \dodoi{10.1086/382068}

\bibitem[{{Lawler} {et~al.}(2015){Lawler}, {Sneden}, \& {Cowan}}]{lawler15}
---. 2015, \apjs, 220, 13, \dodoi{10.1088/0067-0049/220/1/13}

\bibitem[{{Lawler} {et~al.}(2009){Lawler}, {Sneden}, {Cowan}, {Ivans}, \& {Den
  Hartog}}]{lawler09}
{Lawler}, J.~E., {Sneden}, C., {Cowan}, J.~J., {Ivans}, I.~I., \& {Den Hartog},
  E.~A. 2009, \apjs, 182, 51, \dodoi{10.1088/0067-0049/182/1/51}

\bibitem[{{Lawler} {et~al.}(2008){Lawler}, {Sneden}, {Cowan}, {Wyart}, {Ivans},
  {Sobeck}, {Stockett}, \& {Den Hartog}}]{lawler08}
{Lawler}, J.~E., {Sneden}, C., {Cowan}, J.~J., {et~al.} 2008, \apjs, 178, 71,
  \dodoi{10.1086/589834}

\bibitem[{{Lawler} {et~al.}(2017){Lawler}, {Sneden}, {Nave}, {Den Hartog},
  {Emraho{\u g}lu}, \& {Cowan}}]{lawler17}
{Lawler}, J.~E., {Sneden}, C., {Nave}, G., {et~al.} 2017, \apjs, 228, 10,
  \dodoi{10.3847/1538-4365/228/1/10}

\bibitem[{{Lawler} {et~al.}(2001{\natexlab{b}}){Lawler}, {Wickliffe}, {Cowley},
  \& {Sneden}}]{lawler01tb}
{Lawler}, J.~E., {Wickliffe}, M.~E., {Cowley}, C.~R., \& {Sneden}, C.
  2001{\natexlab{b}}, \apjs, 137, 341, \dodoi{10.1086/323001}

\bibitem[{{Lawler} {et~al.}(2001{\natexlab{c}}){Lawler}, {Wickliffe}, {den
  Hartog}, \& {Sneden}}]{lawler01eu}
{Lawler}, J.~E., {Wickliffe}, M.~E., {den Hartog}, E.~A., \& {Sneden}, C.
  2001{\natexlab{c}}, \apj, 563, 1075, \dodoi{10.1086/323407}

\bibitem[{{Lawler} {et~al.}(2014){Lawler}, {Wood}, {Den Hartog}, {Feigenson},
  {Sneden}, \& {Cowan}}]{lawler14}
{Lawler}, J.~E., {Wood}, M.~P., {Den Hartog}, E.~A., {et~al.} 2014, \apjs, 215,
  20, \dodoi{10.1088/0067-0049/215/2/20}

\bibitem[{{Lawler} {et~al.}(2001{\natexlab{d}}){Lawler}, {Wyart}, \&
  {Blaise}}]{lawler01tbhfs}
{Lawler}, J.~E., {Wyart}, J.-F., \& {Blaise}, J. 2001{\natexlab{d}}, \apjs,
  137, 351, \dodoi{10.1086/323000}

\bibitem[{{Li} {et~al.}(2007){Li}, {Chatelain}, {Holt}, {Rehse}, {Rosner}, \&
  {Scholl}}]{li07}
{Li}, R., {Chatelain}, R., {Holt}, R.~A., {et~al.} 2007, \physscr, 76, 577,
  \dodoi{10.1088/0031-8949/76/5/028}

\bibitem[{{Lind} {et~al.}(2011){Lind}, {Asplund}, {Barklem}, \&
  {Belyaev}}]{lind11}
{Lind}, K., {Asplund}, M., {Barklem}, P.~S., \& {Belyaev}, A.~K. 2011, \aap,
  528, A103, \dodoi{10.1051/0004-6361/201016095}

\bibitem[{{Lind} {et~al.}(2012){Lind}, {Bergemann}, \& {Asplund}}]{lind12}
{Lind}, K., {Bergemann}, M., \& {Asplund}, M. 2012, \mnras, 427, 50,
  \dodoi{10.1111/j.1365-2966.2012.21686.x}

\bibitem[{{Lindegren} {et~al.}(2018){Lindegren}, {Hern{\'a}ndez}, {Bombrun},
  {Klioner}, {Bastian}, {Ramos-Lerate}, {de Torres}, {Steidelm{\"u}ller},
  {Stephenson}, {Hobbs}, {Lammers}, {Biermann}, {Geyer}, {Hilger}, {Michalik},
  {Stampa}, {McMillan}, {Casta{\~n}eda}, {Clotet}, {Comoretto}, {Davidson},
  {Fabricius}, {Gracia}, {Hambly}, {Hutton}, {Mora}, {Portell}, {van Leeuwen},
  {Abbas}, {Abreu}, {Altmann}, {Andrei}, {Anglada}, {Balaguer-N{\'u}{\~n}ez},
  {Barache}, {Becciani}, {Bertone}, {Bianchi}, {Bouquillon}, {Bourda},
  {Br{\"u}semeister}, {Bucciarelli}, {Busonero}, {Buzzi}, {Cancelliere},
  {Carlucci}, {Charlot}, {Cheek}, {Crosta}, {Crowley}, {de Bruijne}, {de
  Felice}, {Drimmel}, {Esquej}, {Fienga}, {Fraile}, {Gai}, {Garralda},
  {Gonz{\'a}lez-Vidal}, {Guerra}, {Hauser}, {Hofmann}, {Holl}, {Jordan},
  {Lattanzi}, {Lenhardt}, {Liao}, {Licata}, {Lister}, {L{\"o}ffler},
  {Marchant}, {Martin-Fleitas}, {Messineo}, {Mignard}, {Morbidelli}, {Poggio},
  {Riva}, {Rowell}, {Salguero}, {Sarasso}, {Sciacca}, {Siddiqui}, {Smart},
  {Spagna}, {Steele}, {Taris}, {Torra}, {van Elteren}, {van Reeven}, \&
  {Vecchiato}}]{lindegren18}
{Lindegren}, L., {Hern{\'a}ndez}, J., {Bombrun}, A., {et~al.} 2018, \aap, 616,
  A2, \dodoi{10.1051/0004-6361/201832727}

\bibitem[{{Liu} {et~al.}(2014){Liu}, {Deng}, {Carlin}, {Smith}, {Li},
  {Newberg}, {Gao}, {Yang}, {Xue}, {Xu}, {Zhang}, {Xin}, {Wu}, \&
  {Jin}}]{liu14}
{Liu}, C., {Deng}, L.-C., {Carlin}, J.~L., {et~al.} 2014, \apj, 790, 110,
  \dodoi{10.1088/0004-637X/790/2/110}

\bibitem[{{Ljung} {et~al.}(2006){Ljung}, {Nilsson}, {Asplund}, \&
  {Johansson}}]{ljung06}
{Ljung}, G., {Nilsson}, H., {Asplund}, M., \& {Johansson}, S. 2006, \aap, 456,
  1181, \dodoi{10.1051/0004-6361:20065212}

\bibitem[{{Martinez}(2002)}]{martinez02}
{Martinez}, P. 2002, The Observatory, 122, 359

\bibitem[{{Mashonkina} {et~al.}(2012){Mashonkina}, {Ryabtsev}, \&
  {Frebel}}]{mashonkina12}
{Mashonkina}, L., {Ryabtsev}, A., \& {Frebel}, A. 2012, \aap, 540, A98,
  \dodoi{10.1051/0004-6361/201218790}

\bibitem[{{McCall}(2004)}]{mccall04}
{McCall}, M.~L. 2004, \aj, 128, 2144, \dodoi{10.1086/424933}

\bibitem[{{McConnachie}(2012)}]{mcconnachie12}
{McConnachie}, A.~W. 2012, \aj, 144, 4, \dodoi{10.1088/0004-6256/144/1/4}

\bibitem[{{McDonald} {et~al.}(2012){McDonald}, {Zijlstra}, \&
  {Boyer}}]{mcdonald12}
{McDonald}, I., {Zijlstra}, A.~A., \& {Boyer}, M.~L. 2012, \mnras, 427, 343,
  \dodoi{10.1111/j.1365-2966.2012.21873.x}

\bibitem[{{McWilliam}(1998)}]{mcwilliam98}
{McWilliam}, A. 1998, \aj, 115, 1640, \dodoi{10.1086/300289}

\bibitem[{{Morton}(2000)}]{morton00}
{Morton}, D.~C. 2000, \apjs, 130, 403, \dodoi{10.1086/317349}

\bibitem[{{Navarrete} {et~al.}(2015){Navarrete}, {Chanam{\'e}},
  {Ram{\'{\i}}rez}, {Meza}, {Anglada-Escud{\'e}}, \& {Shkolnik}}]{navarette15}
{Navarrete}, C., {Chanam{\'e}}, J., {Ram{\'{\i}}rez}, I., {et~al.} 2015, \apj,
  808, 103, \dodoi{10.1088/0004-637X/808/1/103}

\bibitem[{{Nilsson} \& {Ivarsson}(2008)}]{nilsson08}
{Nilsson}, H., \& {Ivarsson}, S. 2008, \aap, 492, 609,
  \dodoi{10.1051/0004-6361:200811019}

\bibitem[{{Nilsson} {et~al.}(2002{\natexlab{a}}){Nilsson}, {Ivarsson},
  {Johansson}, \& {Lundberg}}]{nilsson02u}
{Nilsson}, H., {Ivarsson}, S., {Johansson}, S., \& {Lundberg}, H.
  2002{\natexlab{a}}, \aap, 381, 1090, \dodoi{10.1051/0004-6361:20011540}

\bibitem[{{Nilsson} {et~al.}(2002{\natexlab{b}}){Nilsson}, {Zhang}, {Lundberg},
  {Johansson}, \& {Nordstr{\"o}m}}]{nilsson02th}
{Nilsson}, H., {Zhang}, Z.~G., {Lundberg}, H., {Johansson}, S., \&
  {Nordstr{\"o}m}, B. 2002{\natexlab{b}}, \aap, 382, 368,
  \dodoi{10.1051/0004-6361:20011597}

\bibitem[{{Norris} {et~al.}(1985){Norris}, {Bessell}, \& {Pickles}}]{norris85}
{Norris}, J., {Bessell}, M.~S., \& {Pickles}, A.~J. 1985, \apjs, 58, 463,
  \dodoi{10.1086/191049}

\bibitem[{{O'Brian} {et~al.}(1991){O'Brian}, {Wickliffe}, {Lawler}, {Whaling},
  \& {Brault}}]{obrian91}
{O'Brian}, T.~R., {Wickliffe}, M.~E., {Lawler}, J.~E., {Whaling}, W., \&
  {Brault}, J.~W. 1991, Journal of the Optical Society of America B Optical
  Physics, 8, 1185, \dodoi{10.1364/JOSAB.8.001185}

\bibitem[{{Pehlivan Rhodin} {et~al.}(2017){Pehlivan Rhodin}, {Hartman},
  {Nilsson}, \& {J{\"o}nsson}}]{pehlivanrhodin17}
{Pehlivan Rhodin}, A., {Hartman}, H., {Nilsson}, H., \& {J{\"o}nsson}, P. 2017,
  \aap, 598, A102, \dodoi{10.1051/0004-6361/201629849}

\bibitem[{{Placco} {et~al.}(2014){Placco}, {Frebel}, {Beers}, \&
  {Stancliffe}}]{placco14c}
{Placco}, V.~M., {Frebel}, A., {Beers}, T.~C., \& {Stancliffe}, R.~J. 2014,
  \apj, 797, 21, \dodoi{10.1088/0004-637X/797/1/21}

\bibitem[{{Placco} {et~al.}(2017){Placco}, {Holmbeck}, {Frebel}, {Beers},
  {Surman}, {Ji}, {Ezzeddine}, {Points}, {Kaleida}, {Hansen}, {Sakari}, \&
  {Casey}}]{placco17rpro}
{Placco}, V.~M., {Holmbeck}, E.~M., {Frebel}, A., {et~al.} 2017, \apj, 844, 18,
  \dodoi{10.3847/1538-4357/aa78ef}

\bibitem[{{Preston} {et~al.}(2006){Preston}, {Sneden}, {Thompson}, {Shectman},
  \& {Burley}}]{preston06}
{Preston}, G.~W., {Sneden}, C., {Thompson}, I.~B., {Shectman}, S.~A., \&
  {Burley}, G.~S. 2006, \aj, 132, 85, \dodoi{10.1086/504425}

\bibitem[{{Qian} \& {Wasserburg}(2007)}]{qian07}
{Qian}, Y.-Z., \& {Wasserburg}, G.~J. 2007, \physrep, 442, 237,
  \dodoi{10.1016/j.physrep.2007.02.006}

\bibitem[{{Quinet} {et~al.}(2006){Quinet}, {Palmeri}, {Bi{\'e}mont},
  {Jorissen}, {van Eck}, {Svanberg}, {Xu}, \& {Plez}}]{quinet06}
{Quinet}, P., {Palmeri}, P., {Bi{\'e}mont}, {\'E}., {et~al.} 2006, \aap, 448,
  1207, \dodoi{10.1051/0004-6361:20053852}

\bibitem[{{R Core Team}(2013)}]{rsoftware}
{R Core Team}. 2013, R: A Language and Environment for Statistical Computing, R
  Foundation for Statistical Computing, Vienna, Austria

\bibitem[{{Ram{\'{\i}}rez} {et~al.}(2007){Ram{\'{\i}}rez}, {Allende Prieto}, \&
  {Lambert}}]{ramirez07}
{Ram{\'{\i}}rez}, I., {Allende Prieto}, C., \& {Lambert}, D.~L. 2007, \aap,
  465, 271, \dodoi{10.1051/0004-6361:20066619}

\bibitem[{{Ram{\'{\i}}rez} \& {Mel{\'e}ndez}(2005)}]{ramirez05b}
{Ram{\'{\i}}rez}, I., \& {Mel{\'e}ndez}, J. 2005, \apj, 626, 465,
  \dodoi{10.1086/430102}

\bibitem[{{Roederer} \& {Barklem}(2018)}]{roederer18a}
{Roederer}, I.~U., \& {Barklem}, P.~S. 2018, \apj, 857, 2,
  \dodoi{10.3847/1538-4357/aab71f}

\bibitem[{{Roederer} {et~al.}(2010{\natexlab{a}}){Roederer}, {Cowan},
  {Karakas}, {Kratz}, {Lugaro}, {Simmerer}, {Farouqi}, \&
  {Sneden}}]{roederer10c}
{Roederer}, I.~U., {Cowan}, J.~J., {Karakas}, A.~I., {et~al.}
  2010{\natexlab{a}}, \apj, 724, 975, \dodoi{10.1088/0004-637X/724/2/975}

\bibitem[{{Roederer} {et~al.}(2018{\natexlab{a}}){Roederer}, {Hattori}, \&
  {Valluri}}]{roederer18c}
{Roederer}, I.~U., {Hattori}, K., \& {Valluri}, M. 2018{\natexlab{a}}, \aj, in
  press

\bibitem[{{Roederer} \& {Lawler}(2012)}]{roederer12b}
{Roederer}, I.~U., \& {Lawler}, J.~E. 2012, \apj, 750, 76,
  \dodoi{10.1088/0004-637X/750/1/76}

\bibitem[{{Roederer} {et~al.}(2008){Roederer}, {Lawler}, {Sneden}, {Cowan},
  {Sobeck}, \& {Pilachowski}}]{roederer08a}
{Roederer}, I.~U., {Lawler}, J.~E., {Sneden}, C., {et~al.} 2008, \apj, 675,
  723, \dodoi{10.1086/526452}

\bibitem[{{Roederer} {et~al.}(2016){Roederer}, {Mateo}, {Bailey}, {Spencer},
  {Crane}, \& {Shectman}}]{roederer16a}
{Roederer}, I.~U., {Mateo}, M., {Bailey}, J.~I., {et~al.} 2016, \mnras, 455,
  2417, \dodoi{10.1093/mnras/stv2462}

\bibitem[{{Roederer} {et~al.}(2014{\natexlab{a}}){Roederer}, {Preston},
  {Thompson}, {Shectman}, {Sneden}, {Burley}, \& {Kelson}}]{roederer14c}
{Roederer}, I.~U., {Preston}, G.~W., {Thompson}, I.~B., {et~al.}
  2014{\natexlab{a}}, \aj, 147, 136, \dodoi{10.1088/0004-6256/147/6/136}

\bibitem[{{Roederer} {et~al.}(2010{\natexlab{b}}){Roederer}, {Sneden},
  {Lawler}, \& {Cowan}}]{roederer10b}
{Roederer}, I.~U., {Sneden}, C., {Lawler}, J.~E., \& {Cowan}, J.~J.
  2010{\natexlab{b}}, \apjl, 714, L123, \dodoi{10.1088/2041-8205/714/1/L123}

\bibitem[{{Roederer} {et~al.}(2018{\natexlab{b}}){Roederer}, {Sneden},
  {Lawler}, {Sobeck}, {Cowan}, \& {Boesgaard}}]{roederer18b}
{Roederer}, I.~U., {Sneden}, C., {Lawler}, J.~E., {et~al.} 2018{\natexlab{b}},
  \apj, 860, 125, \dodoi{10.3847/1538-4357/aac6df}

\bibitem[{{Roederer} \& {Thompson}(2015)}]{roederer15}
{Roederer}, I.~U., \& {Thompson}, I.~B. 2015, \mnras, 449, 3889,
  \dodoi{10.1093/mnras/stv546}

\bibitem[{{Roederer} {et~al.}(2012){Roederer}, {Lawler}, {Sobeck}, {Beers},
  {Cowan}, {Frebel}, {Ivans}, {Schatz}, {Sneden}, \& {Thompson}}]{roederer12d}
{Roederer}, I.~U., {Lawler}, J.~E., {Sobeck}, J.~S., {et~al.} 2012, \apjs, 203,
  27, \dodoi{10.1088/0067-0049/203/2/27}

\bibitem[{{Roederer} {et~al.}(2014{\natexlab{b}}){Roederer}, {Schatz},
  {Lawler}, {Beers}, {Cowan}, {Frebel}, {Ivans}, {Sneden}, \&
  {Sobeck}}]{roederer14d}
{Roederer}, I.~U., {Schatz}, H., {Lawler}, J.~E., {et~al.} 2014{\natexlab{b}},
  \apj, 791, 32, \dodoi{10.1088/0004-637X/791/1/32}

\bibitem[{{Ruffoni} {et~al.}(2014){Ruffoni}, {Den Hartog}, {Lawler}, {Brewer},
  {Lind}, {Nave}, \& {Pickering}}]{ruffoni14}
{Ruffoni}, M.~P., {Den Hartog}, E.~A., {Lawler}, J.~E., {et~al.} 2014, \mnras,
  441, 3127, \dodoi{10.1093/mnras/stu780}

\bibitem[{{Sadakane} {et~al.}(2004){Sadakane}, {Arimoto}, {Ikuta}, {Aoki},
  {Jablonka}, \& {Tajitsu}}]{sadakane04}
{Sadakane}, K., {Arimoto}, N., {Ikuta}, C., {et~al.} 2004, \pasj, 56, 1041,
  \dodoi{10.1093/pasj/56.6.1041}

\bibitem[{{Sakari} {et~al.}(2018{\natexlab{a}}){Sakari}, {Placco}, {Farrell},
  {Wallerstein}, {Beers}, {Frebel}, {Hansen}, {Holmbeck}, {Roederer}, {Sneden},
  {Venn}, {Davis}, {Matijevic}, \& {Wyse}}]{sakari18north}
{Sakari}, C.~M., {Placco}, V.~M., {Farrell}, E.~M., {et~al.}
  2018{\natexlab{a}}, AAS Journals, submitted

\bibitem[{{Sakari} {et~al.}(2018{\natexlab{b}}){Sakari}, {Placco}, {Hansen},
  {Holmbeck}, {Beers}, {Frebel}, {Roederer}, {Venn}, {Wallerstein}, {Davis},
  {Farrell}, \& {Yong}}]{sakari18a}
{Sakari}, C.~M., {Placco}, V.~M., {Hansen}, T., {et~al.} 2018{\natexlab{b}},
  \apjl, 854, L20, \dodoi{10.3847/2041-8213/aaa9b4}

\bibitem[{{Schatz} {et~al.}(2002){Schatz}, {Toenjes}, {Pfeiffer}, {Beers},
  {Cowan}, {Hill}, \& {Kratz}}]{schatz02}
{Schatz}, H., {Toenjes}, R., {Pfeiffer}, B., {et~al.} 2002, \apj, 579, 626,
  \dodoi{10.1086/342939}

\bibitem[{{Schlafly} \& {Finkbeiner}(2011)}]{schlafly11}
{Schlafly}, E.~F., \& {Finkbeiner}, D.~P. 2011, \apj, 737, 103,
  \dodoi{10.1088/0004-637X/737/2/103}

\bibitem[{{Schlaufman} \& {Casey}(2014)}]{schlaufman14}
{Schlaufman}, K.~C., \& {Casey}, A.~R. 2014, \apj, 797, 13,
  \dodoi{10.1088/0004-637X/797/1/13}

\bibitem[{{Shetrone} {et~al.}(2001){Shetrone}, {C{\^o}t{\'e}}, \&
  {Sargent}}]{shetrone01}
{Shetrone}, M.~D., {C{\^o}t{\'e}}, P., \& {Sargent}, W.~L.~W. 2001, \apj, 548,
  592, \dodoi{10.1086/319022}

\bibitem[{{Siqueira Mello} {et~al.}(2013){Siqueira Mello}, {Spite}, {Barbuy},
  {Spite}, {Caffau}, {Hill}, {Wanajo}, {Primas}, {Plez}, {Cayrel}, {Andersen},
  {Nordstr{\"o}m}, {Sneden}, {Beers}, {Bonifacio}, {Fran{\c c}ois}, \&
  {Molaro}}]{siqueiramello13}
{Siqueira Mello}, C., {Spite}, M., {Barbuy}, B., {et~al.} 2013, \aap, 550,
  A122, \dodoi{10.1051/0004-6361/201219949}

\bibitem[{{Smith} {et~al.}(1998){Smith}, {Lambert}, \& {Nissen}}]{smith98}
{Smith}, V.~V., {Lambert}, D.~L., \& {Nissen}, P.~E. 1998, \apj, 506, 405,
  \dodoi{10.1086/306238}

\bibitem[{{Sneden} {et~al.}(2008){Sneden}, {Cowan}, \& {Gallino}}]{sneden08}
{Sneden}, C., {Cowan}, J.~J., \& {Gallino}, R. 2008, \araa, 46, 241,
  \dodoi{10.1146/annurev.astro.46.060407.145207}

\bibitem[{{Sneden} {et~al.}(2016){Sneden}, {Cowan}, {Kobayashi}, {Pignatari},
  {Lawler}, {Den Hartog}, \& {Wood}}]{sneden16}
{Sneden}, C., {Cowan}, J.~J., {Kobayashi}, C., {et~al.} 2016, \apj, 817, 53,
  \dodoi{10.3847/0004-637X/817/1/53}

\bibitem[{{Sneden} {et~al.}(2009){Sneden}, {Lawler}, {Cowan}, {Ivans}, \& {Den
  Hartog}}]{sneden09}
{Sneden}, C., {Lawler}, J.~E., {Cowan}, J.~J., {Ivans}, I.~I., \& {Den Hartog},
  E.~A. 2009, \apjs, 182, 80, \dodoi{10.1088/0067-0049/182/1/80}

\bibitem[{{Sneden} {et~al.}(1994){Sneden}, {Preston}, {McWilliam}, \&
  {Searle}}]{sneden94}
{Sneden}, C., {Preston}, G.~W., {McWilliam}, A., \& {Searle}, L. 1994, \apjl,
  431, L27, \dodoi{10.1086/187464}

\bibitem[{{Sneden} {et~al.}(2003){Sneden}, {Cowan}, {Lawler}, {Ivans},
  {Burles}, {Beers}, {Primas}, {Hill}, {Truran}, {Fuller}, {Pfeiffer}, \&
  {Kratz}}]{sneden03a}
{Sneden}, C., {Cowan}, J.~J., {Lawler}, J.~E., {et~al.} 2003, \apj, 591, 936,
  \dodoi{10.1086/375491}

\bibitem[{{Sneden}(1973)}]{sneden73}
{Sneden}, C.~A. 1973, PhD thesis, The University of Texas at Austin.

\bibitem[{{Sobeck} {et~al.}(2007){Sobeck}, {Lawler}, \& {Sneden}}]{sobeck07}
{Sobeck}, J.~S., {Lawler}, J.~E., \& {Sneden}, C. 2007, \apj, 667, 1267,
  \dodoi{10.1086/519987}

\bibitem[{{Sobeck} {et~al.}(2011){Sobeck}, {Kraft}, {Sneden}, {Preston},
  {Cowan}, {Smith}, {Thompson}, {Shectman}, \& {Burley}}]{sobeck11}
{Sobeck}, J.~S., {Kraft}, R.~P., {Sneden}, C., {et~al.} 2011, \aj, 141, 175,
  \dodoi{10.1088/0004-6256/141/6/175}

\bibitem[{{Takeda} {et~al.}(2002){Takeda}, {Zhao}, {Chen}, {Qiu}, \&
  {Takada-Hidai}}]{takeda02}
{Takeda}, Y., {Zhao}, G., {Chen}, Y.-Q., {Qiu}, H.-M., \& {Takada-Hidai}, M.
  2002, \pasj, 54, 275, \dodoi{10.1093/pasj/54.2.275}

\bibitem[{{Tanvir} {et~al.}(2017){Tanvir}, {Levan},
  {Gonz{\'a}lez-Fern{\'a}ndez}, {Korobkin}, {Mandel}, {Rosswog}, {Hjorth},
  {D'Avanzo}, {Fruchter}, {Fryer}, {Kangas}, {Milvang-Jensen}, {Rosetti},
  {Steeghs}, {Wollaeger}, {Cano}, {Copperwheat}, {Covino}, {D'Elia}, {de Ugarte
  Postigo}, {Evans}, {Even}, {Fairhurst}, {Figuera Jaimes}, {Fontes}, {Fujii},
  {Fynbo}, {Gompertz}, {Greiner}, {Hodosan}, {Irwin}, {Jakobsson},
  {J{\o}rgensen}, {Kann}, {Lyman}, {Malesani}, {McMahon}, {Melandri},
  {O'Brien}, {Osborne}, {Palazzi}, {Perley}, {Pian}, {Piranomonte}, {Rabus},
  {Rol}, {Rowlinson}, {Schulze}, {Sutton}, {Th{\"o}ne}, {Ulaczyk}, {Watson},
  {Wiersema}, \& {Wijers}}]{tanvir17}
{Tanvir}, N.~R., {Levan}, A.~J., {Gonz{\'a}lez-Fern{\'a}ndez}, C., {et~al.}
  2017, \apjl, 848, L27, \dodoi{10.3847/2041-8213/aa90b6}

\bibitem[{{Thielemann} {et~al.}(2017){Thielemann}, {Eichler}, {Panov}, \&
  {Wehmeyer}}]{thielemann17}
{Thielemann}, F.-K., {Eichler}, M., {Panov}, I.~V., \& {Wehmeyer}, B. 2017,
  Annual Review of Nuclear and Particle Science, 67, 253,
  \dodoi{10.1146/annurev-nucl-101916-123246}

\bibitem[{{Tody}(1993)}]{tody93}
{Tody}, D. 1993, in Astronomical Society of the Pacific Conference Series,
  Vol.~52, Astronomical Data Analysis Software and Systems II, ed. R.~J.
  {Hanisch}, R.~J.~V. {Brissenden}, \& J.~{Barnes}, 173

\bibitem[{{Tsujimoto} {et~al.}(2017){Tsujimoto}, {Matsuno}, {Aoki}, {Ishigaki},
  \& {Shigeyama}}]{tsujimoto17}
{Tsujimoto}, T., {Matsuno}, T., {Aoki}, W., {Ishigaki}, M.~N., \& {Shigeyama},
  T. 2017, \apjl, 850, L12, \dodoi{10.3847/2041-8213/aa9886}

\bibitem[{{Unsold}(1955)}]{unsold55}
{Unsold}, A. 1955, {Physik der Sternatmospharen, MIT besonderer
  Berucksichtigung der Sonne.} (Berlin, Springer)

\bibitem[{{van der Walt} {et~al.}(2011){van der Walt}, {Colbert}, \&
  {Varoquaux}}]{vanderwalt11}
{van der Walt}, S., {Colbert}, S.~C., \& {Varoquaux}, G. 2011, Computing in
  Science Engineering, 13, 22, \dodoi{10.1109/MCSE.2011.37}

\bibitem[{{Van Eck} {et~al.}(2003){Van Eck}, {Goriely}, {Jorissen}, \&
  {Plez}}]{vaneck03}
{Van Eck}, S., {Goriely}, S., {Jorissen}, A., \& {Plez}, B. 2003, \aap, 404,
  291, \dodoi{10.1051/0004-6361:20030447}

\bibitem[{{Walker} {et~al.}(2016){Walker}, {Mateo}, {Olszewski}, {Koposov},
  {Belokurov}, {Jethwa}, {Nidever}, {Bonnivard}, {Bailey}, {Bell}, \&
  {Loebman}}]{walker16}
{Walker}, M.~G., {Mateo}, M., {Olszewski}, E.~W., {et~al.} 2016, \apj, 819, 53,
  \dodoi{10.3847/0004-637X/819/1/53}

\bibitem[{{Wickliffe} \& {Lawler}(1997)}]{wickliffe97tm}
{Wickliffe}, M.~E., \& {Lawler}, J.~E. 1997, Journal of the Optical Society of
  America B Optical Physics, 14, 737, \dodoi{10.1364/JOSAB.14.000737}

\bibitem[{{Wickliffe} {et~al.}(2000){Wickliffe}, {Lawler}, \&
  {Nave}}]{wickliffe00}
{Wickliffe}, M.~E., {Lawler}, J.~E., \& {Nave}, G. 2000, \jqsrt, 66, 363,
  \dodoi{10.1016/S0022-4073(99)00173-9}

\bibitem[{{Wickliffe} {et~al.}(1994){Wickliffe}, {Salih}, \&
  {Lawler}}]{wickliffe94}
{Wickliffe}, M.~E., {Salih}, S., \& {Lawler}, J.~E. 1994, \jqsrt, 51, 545,
  \dodoi{10.1016/0022-4073(94)90108-2}

\bibitem[{{Wood} {et~al.}(2014{\natexlab{a}}){Wood}, {Lawler}, {Den Hartog},
  {Sneden}, \& {Cowan}}]{wood14v}
{Wood}, M.~P., {Lawler}, J.~E., {Den Hartog}, E.~A., {Sneden}, C., \& {Cowan},
  J.~J. 2014{\natexlab{a}}, \apjs, 214, 18, \dodoi{10.1088/0067-0049/214/2/18}

\bibitem[{{Wood} {et~al.}(2013){Wood}, {Lawler}, {Sneden}, \& {Cowan}}]{wood13}
{Wood}, M.~P., {Lawler}, J.~E., {Sneden}, C., \& {Cowan}, J.~J. 2013, \apjs,
  208, 27, \dodoi{10.1088/0067-0049/208/2/27}

\bibitem[{{Wood} {et~al.}(2014{\natexlab{b}}){Wood}, {Lawler}, {Sneden}, \&
  {Cowan}}]{wood14ni}
---. 2014{\natexlab{b}}, \apjs, 211, 20, \dodoi{10.1088/0067-0049/211/2/20}

\bibitem[{{Xu} {et~al.}(2007){Xu}, {Svanberg}, {Quinet}, {Palmeri}, \&
  {Bi{\'e}mont}}]{xu07}
{Xu}, H.~L., {Svanberg}, S., {Quinet}, P., {Palmeri}, P., \& {Bi{\'e}mont},
  {\'E}. 2007, \jqsrt, 104, 52, \dodoi{10.1016/j.jqsrt.2006.08.010}

\end{thebibliography}

\end{document}